\documentclass[reprint,amsmath,amssymb,aps]{revtex4-2}
\pdfoutput=1
\usepackage{mathtools}
\usepackage{graphicx}
\usepackage{dcolumn}
\usepackage{bm}
\usepackage{braket}
\usepackage{siunitx}
\usepackage{enumerate}
\usepackage[lined,boxruled]{algorithm2e}
\usepackage{bbm}
\usepackage{booktabs}
\usepackage{comment}

\usepackage[utf8]{inputenc}
\usepackage[english]{babel}

\usepackage{amsthm}
\newcommand{\secref}[1]{{\rm Section~\ref{#1}}}
\newcommand{\figref}[1]{{\rm Figure~\ref{#1}}}
\newcommand{\tabref}[1]{{\rm Table~\ref{#1}}}
\newcommand{\meqref}[1]{{\rm Equation~\eqref{#1}}}

\newcommand{\F}{\mathcal{F}}

\DeclareMathOperator{\Tr}{Tr}

\begin{document}

\preprint{APS/123-QED}

\title{Term Grouping and Travelling Salesperson for Digital Quantum Simulation}

\author{Kaiwen Gui$^1$}
\thanks{These two authors contributed equally}
\author{Teague Tomesh$^2$}
\thanks{These two authors contributed equally}
\author{Pranav Gokhale$^3$}
\author{Yunong Shi$^4$}
\author{Frederic T. Chong$^3$}
\author{Margaret Martonosi$^2$}
\author{Martin Suchara$^{1,5}$}
\affiliation{$^1$Pritzker School of Molecular Engineering, University of Chicago}
\affiliation{$^2$Department of Computer Science, Princeton University}
\affiliation{$^3$Department of Computer Science, University of Chicago}
\affiliation{$^4$Department of Physics, University of Chicago}
\affiliation{$^5$Mathematics and Computer Science Division, Argonne National Laboratory}


\date{\today}


\begin{abstract}
Digital simulation of quantum dynamics by evaluating the time evolution of a Hamiltonian is the initially proposed application of quantum computing.  The large number of quantum gates required for emulating the complete second quantization form of the Hamiltonian, however, makes such an approach unsuitable for near-term devices with limited gate fidelities that cause high physical errors. In addition, Trotter error caused by noncommuting terms can accumulate and harm the overall circuit fidelity, thus causing algorithmic errors. In this paper, we propose a new term ordering strategy, \textit{max-commute-tsp (MCTSP)}, that simultaneously mitigates both algorithmic and physical errors. First, we improve the Trotter fidelity compared with previously proposed optimization by reordering Pauli terms and partitioning them into commuting families. We demonstrate the practicality of this method by constructing and evaluating quantum circuits that simulate different molecular Hamiltonians, together with theoretical explanations for the fidelity improvements from our term grouping method. Second, we describe a new gate cancellation technique that reduces the high gate counts by formulating the gate cancellation problem as a travelling salesperson problem, together with benchmarking experiments. Finally, we also provide benchmarking results that demonstrate the combined advantage of \textit{max-commute-tsp} to mitigate both physical and algorithmic errors via quantum circuit simulation under realistic noise models.
\end{abstract}
\maketitle
\section{Introduction} \label{sec:introduction}
The idea of simulating unknown quantum systems by using controllable quantum systems was originally proposed by Richard Feynman in 1982 \cite{Feynman1982} and has remained one of the major motivations for building quantum computers. Similar to some other quantum algorithms, it can provide exponential speed-ups compared with classical simulation of quantum systems \cite{Lloyd1996simulators}. 

Digital quantum simulation (DQS) aims to map a quantum evolution, defined by a time-dependent Hamiltonian, to a digital quantum circuit. The digital quantum circuit is executed with respect to some chosen evolution time in order to ``emulate" the real Hamiltonian evolution. Typically, one will initialize the Hamiltonian using the second quantization form and convert to the qubit representation using mapping methods such as Jordan-Wigner~\cite{jordan1928pauli} or Bravyi-Kitaev~\cite{seeley2012bravyi}. Then the resulting Pauli terms are mapped to the digital quantum circuit (more details are given in Section~\ref{sec:background}).

The na\"ive presentations of the quantum dynamics algorithm have extremely high depth (e.g., $10^7$ gates \cite{tranter2018comparison}), thus leading to high physical errors. This is due to the fact that Pauli terms need to be concatenated in the same quantum circuit (details in Section~\ref{sec:circuit_construction}). This high circuit depth is unsuitable for near-term quantum devices with limited coherence times of hardware qubits\cite{Preskill2018quantumcomputingin} as well as future devices that will consist of qubits with improved but non-negligible error properties.

In addition to the large gate depth, DQS circuits also suffer from large algorithmic errors. In general, the Pauli terms do not commute with each other. Therefore, we can only map the Hamiltonian into an approximated circuit by using the Trotter-Suzuki decomposition \cite{suzuki1990fractal}. This will lead to Trotter errors that can devastate the total circuit fidelity even under perfect quantum gates \cite{Lloyd1996simulators} (details in Section \ref{sec:trotterization}). 
\begin{figure}[h]
    \centering
    \includegraphics[width=0.47\textwidth]{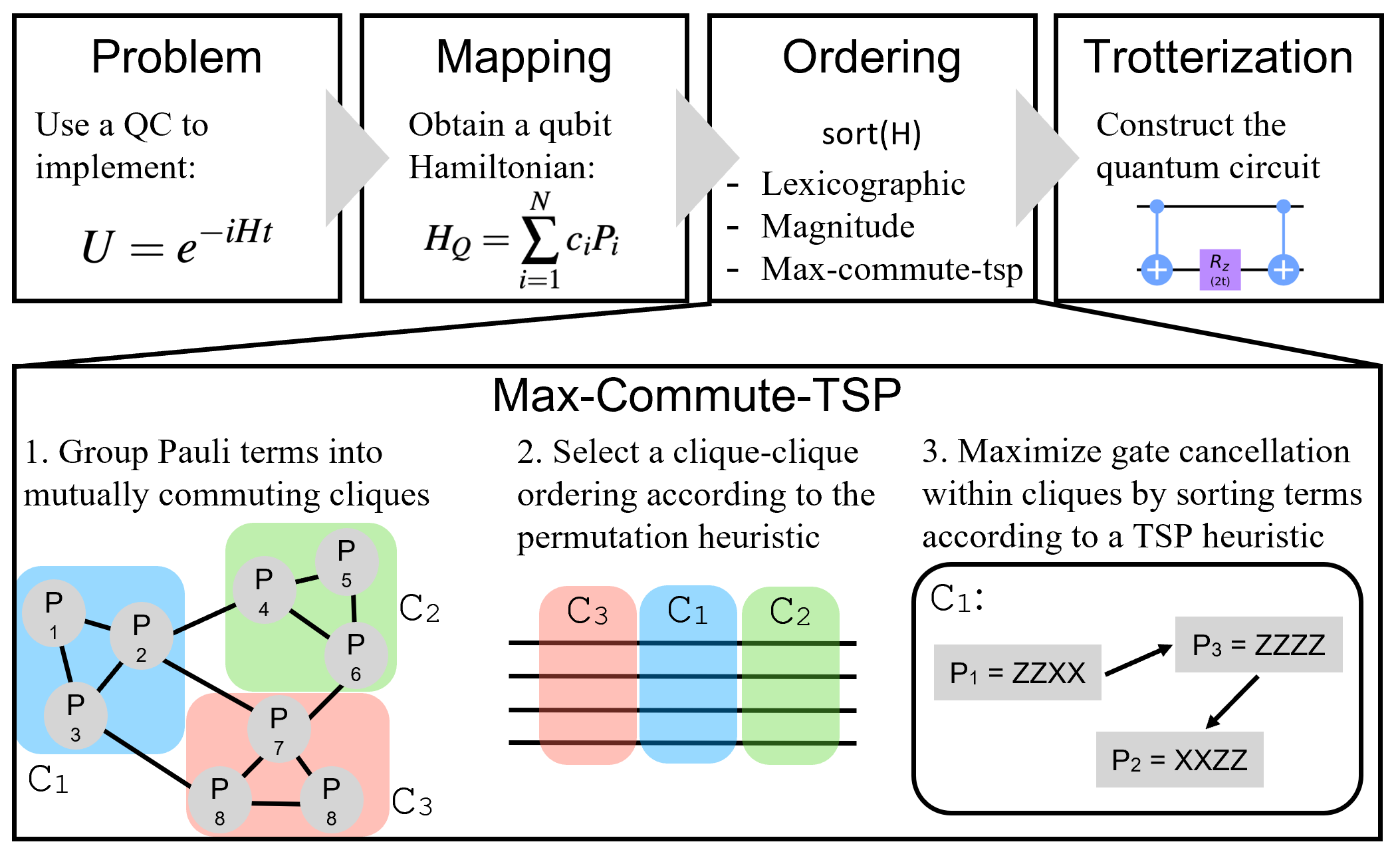}
    \caption{A summary of the DQS compilation process and the \textit{max-commute-tsp} ordering strategy}
    \label{fig:big_pic}
\end{figure}

In this work, we propose a new term-ordering strategy, \textit{max-commute-tsp}, that \textbf{simultaneously} mitigates both \textbf{physical errors} (via gate cancellation) and \textbf{algorithmic} errors (via Trotter error reduction). Figure 1
illustrates the key steps of our method:
\begin{enumerate}
    \item We first group the Pauli terms into commuting groups that aim to provide better Trotter fidelity.
    \item Then we optimize the clique-clique ordering to further boost the Trotter fidelity.
    \item Next, we optimize the ordering of Pauli terms inside each commuting group to achieve maximum gate cancellation. We can reorder the terms inside each commuting group because reordering these Pauli terms will not harm the Trotter fidelity (Corollary 2 in Section \ref{sec:pairwise_gate_cancellation}).
\end{enumerate}

The contributions of this work are as follows:
\begin{itemize}
    \item We provide theoretical explanations on algorithmic (Trotter) error mitigation (Section~\ref{sec:clique_ordering_theory}. This contribution partially overlaps with some concurrent work \cite{tranter2019ordering, Child2019trotter}. Additionally, We analyze the impact of clique-clique ordering on the total circuit fidelity for two and more commuting groups. We also design (Section~\ref{sec:clique_ordering_abstract}) and benchmark (Section~\ref{sec:clique_ordering_experimental}) a polynomial-time clique-clique ordering heuristics that further boost the total circuit Trotter fidelity.
    
    \item We use the traveling salesperson problem (TSP) for gate cancellation abstraction, with theoretical analysis and initial implementations (Section~\ref{sec:pairwise_gate_cancellation} and \ref{sec:gate_cancellation_implementation}).
    
    \item We provide simulation results on separate (Section~\ref{sec:clique_ordering_experimental}, \ref{sec:gate_cancellation_implementation}) and combined (Section~\ref{sec:combined_benchmarking}) effects of physical and algorithmic error mitigation with and without noise models, together with a general, open-source implementation of DQS which can be of use to the quantum computing community as a challenging and practical benchmark.
\end{itemize}

Although we are using the same term-finding technique that was described in our earlier work that optimizes VQE circuits \cite{gokhale2019minimizing}, here it serves a different purpose. In our earlier work, grouping the Pauli terms into cliques was done in order to reduce the total number of required measurements. In this work, we utilize the clique grouping to provide better quantum dynamics circuit fidelity and a more suitable platform for gate cancellation using TSP.

\begin{figure}[h]
    \centering
    \includegraphics[width=1.0\columnwidth]{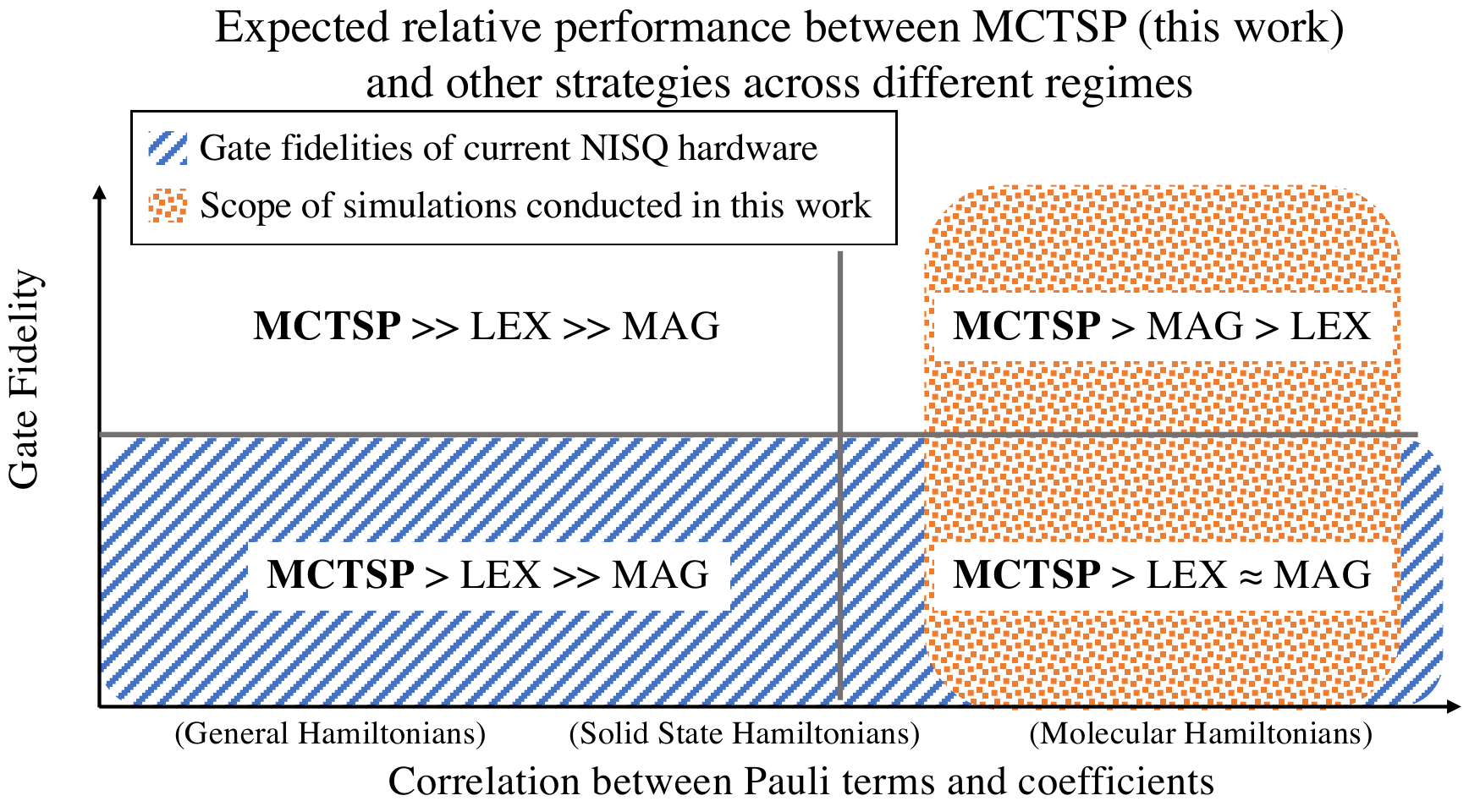}
    \caption{We study the impact of term ordering on HS performance within the regime of molecular Hamiltonians. We expect \textit{max-commute-tsp} (MCTSP) to show good performance compared to previous strategies such as \textit{lexicographic} ordering (LEX) and \textit{magnitude} ordering (MAG) across domains due to its flexibity and ability to mitigate both algorithmic and physical errors.}
    \label{fig:scope}
\end{figure}

In the benchmarking experiments, we only consider simulation and experiments for molecular Hamiltonians because they are a relevant and important application and also easily obtainable via the NIST Chemistry WebBook~\cite{nist1997chemistry} and OpenFermion software package~\cite{mcclean2017openfermion}.
Our strategy is expected to show much better performances for solid state and general Hamiltonians, as illustrated in Fig~\ref{fig:scope}.
\section{Background} \label{sec:background}
The simulation of quantum systems by using programmable quantum computers is one of the most promising applications of quantum computing. Understanding the evolution of a quantum state under some Hamiltonian, H,
\begin{equation}\label{eq:evolution}
    \ket{\psi(t)} = e^{-iHt/{\hbar}}\ket{\psi(0)}
\end{equation}
is a problem of great interest in chemistry and physics.
Classically, the molecular Hamiltonian for electronic structure problems can be exactly simulated by using full configuration interaction with a runtime that scales as a factorial of the number of basis functions~\cite{szabo2012modern}. The same problem can be solved in polynomial time on a quantum computer by leveraging quantum phenomena~\cite{whitfield2011simulation}.

In general, simulation of a molecular system consists of three integrated steps: state preparation, time evolution, and measurement of observables \cite{whitfield2011simulation}. 

In this paper, we focus on the second step: proposing techniques for improving the fidelity and reducing the depth of the quantum circuits used to implement the evolution unitary (Eq.~\eqref{eq:evolution}). We demonstrate the improved fidelity by examining the simulated quantum operation matrix, explained in Section \ref{subsec:fidelity_method}.

\subsection{Quantum Gate Basics} \label{sec:gate_basics}
In DQS circuits, the following quantum gates (represented by their unitary matrices) are used, shown in table~\ref{tab:gate_matrices}:
\begin{table}[h]
    \centering
    \begin{tabular}{ c c }
      \toprule
        \textbf{Gate} & \textbf{Unitary}\\
      \midrule\\
        \addlinespace[-2ex]
        Hadamard & $H=\frac{1}{\sqrt{2}}\begin{pmatrix} 1 & 1 \\ 1 & -1 \end{pmatrix}$ \\
        \addlinespace[1.5ex]
        Z-Rotation & $Rz(\theta)=\begin{pmatrix} e^{-i\frac{\theta}{2}} & 0 \\ 0 & e^{i\frac{\theta}{2}} \end{pmatrix}$ \\
        \addlinespace[1.5ex]
        Phase & $S=\begin{pmatrix} 1 & 0 \\ 0 & i \end{pmatrix}$ \\
        \addlinespace[1.5ex]
        Controlled-Not & $CNOT=\begin{pmatrix} 1 & 0 & 0 & 0 \\ 0 & 1 & 0 & 0 \\ 0 & 0 & 0 & 1 \\ 0 & 0 & 1 & 0 \end{pmatrix}$ \\
      \bottomrule
    \end{tabular}
\caption{Quantum Gates in DQS circuits}
\label{tab:gate_matrices}
\end{table}

\subsection{Quantum Dynamics Circuit}
In general, the quantum evolution unitary $e^{-iHt/\hbar}$ is complicated and hard to implement by using quantum circuits. Fortunately, in  electronic structure problems, there is a relatively easy way~\cite{fermionic_construction_1, fermionic_construction_2, fermionic_construction_3} to implement the summation of the Pauli terms that come after the Jordan-Wigner or Bravyi Kitaev transformations. 

For general electronic structure problems, the molecular Hamiltonians are initialized with the form
\begin{equation}
    H = \sum_{p=1}^N \sum_{q=1}^N h_{pq} a_p^{\dagger} a_q + \sum_{p=1}^N \sum_{q=1}^N \sum_{r=1}^N \sum_{s=1}^N h_{pqrs} a^{\dagger}_p a^{\dagger}_q a_r a_s .
\label{eq:second_quantization}
\end{equation}

We first prepare the initial imperfect coefficients $h_{pq}$ and $h_{pqrs}$ using classical algorithms such as Hartree-Fock. We use OpenFermion \cite{mcclean2017openfermion} to obtain these coefficients.

Next, we map the fermionic operators $a_{p}^{\dagger}$ and $a_{q}$ to the qubit representation. This process can be done via transformation methods such as Jordan-Wigner or Bravyi-Kitaev \cite{mcardle2018quantum}. We denote the resulting strings of the form $\{X, Y, Z, I\}^{\otimes n}$ as ``Pauli terms.'' We also used OpenFermion to obtain these Pauli terms~\cite{mcclean2017openfermion}.

The Pauli terms are then mapped to the quantum circuits.
Consider a simple simulation of the unitary evolution $U = e^{-iZt}$ (Z is the single Pauli Z matrix). It can be mapped to the single Rz(2t) gate (t is the time parameter) defined in Section~\ref{sec:gate_basics}.

\begin{figure}[h!]
    \centering
    \includegraphics[scale=0.5]{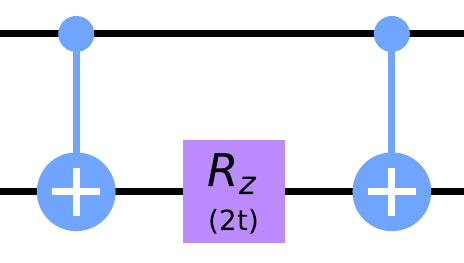}
    \caption{$U = e^{-iZZt}$ circuit}
    \label{fig:circ_ZZ}
\end{figure}

\begin{figure}[h!]
    \centering
    \includegraphics[scale=0.5]{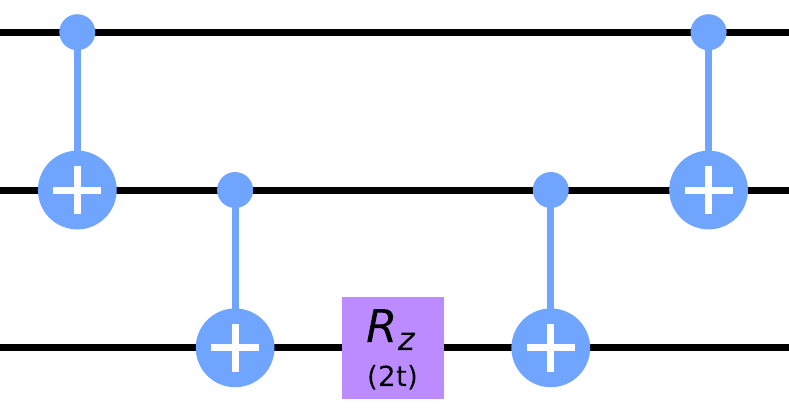}
    \caption{$U = e^{-iZZZt}$ circuit}
    \label{fig:circ_ZZZ}
\end{figure}

Accordingly, the unitary evolution $U = e^{-iZZt}$ and $U = e^{-iZZZt}$ can be evaluated with the circuits in Figs.~\ref{fig:circ_ZZ} and~\ref{fig:circ_ZZZ}.

\begin{figure}[h!]
    \centering
    \includegraphics[scale=0.5]{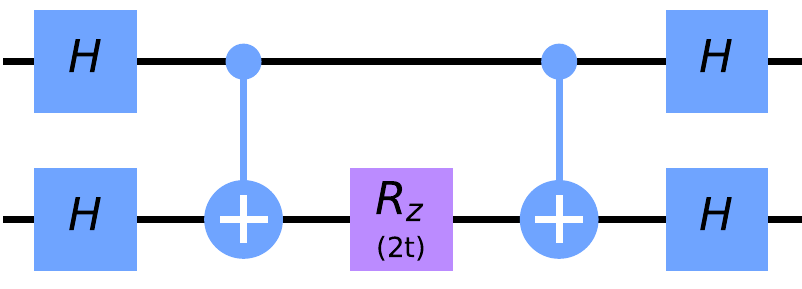}
    \caption{$U = e^{-iXXt}$ circuit}
    \label{fig:circ_XX}
\end{figure}

\begin{figure}[h!]
    \centering
    \includegraphics[scale=0.5]{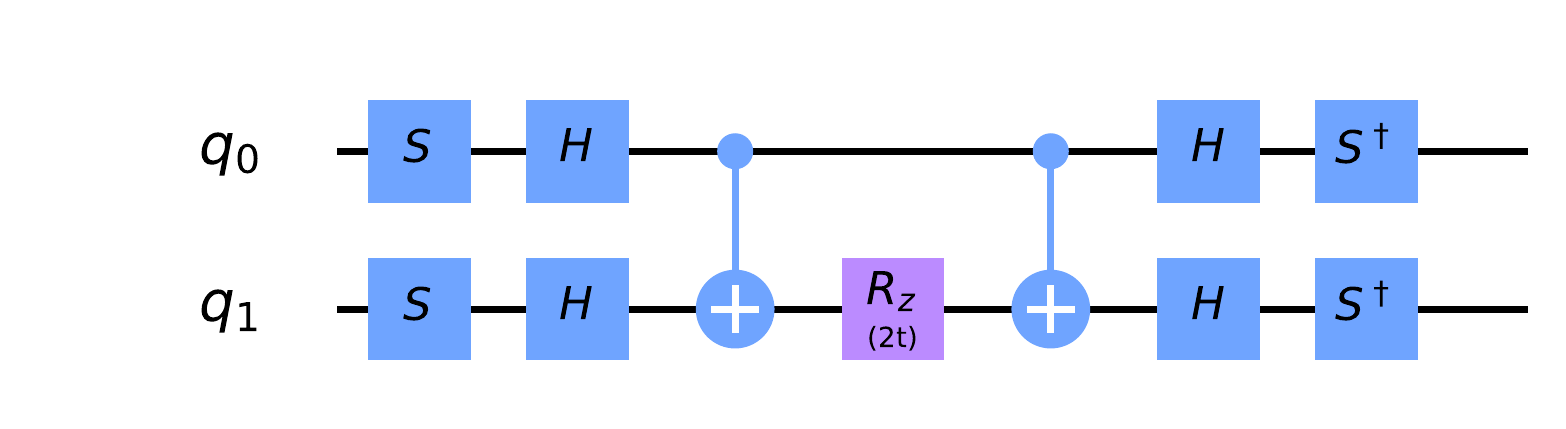}
    \caption{$U = e^{-iYYt}$ circuit}
    \label{fig:circ_YY}
\end{figure}

Similarly, the Pauli X matrix can be simulated by adding Hadamard gates to the front and back of the circuits. Figure~\ref{fig:circ_XX} provides an example of the $U = e^{-i(XX)t}$ evolution.
The Pauli Y matrix can be simulated by adding the Hadamard gate and the S gate in the front and then Hadamard and the $S^{\dagger}$ gate at the end. Figure~\ref{fig:circ_YY} provides an example of the circuit representing $U = e^{-i(YY)t}$.
The coefficients in front of each Pauli term are multiplied with 2t in the Rz gate's parameter.

\subsection{Travelling Salesperson Problem}
The travelling salesperson problem (TSP) aims to solve the following question:
\begin{itemize}
    \item Given a list of cities and the distances between each pair of cities, what is the shortest possible route that visits each city and returns to the origin city?
\end{itemize}

TSP can be modeled as an undirected weighted graph, such that cities are the graph's vertices, paths are the graph's edges, and a path's distance is the edge's weight. It is a minimization problem starting and finishing at a specified vertex after having visited each other vertex exactly once. In another words, it is trying to find a ``Hamiltonian cycle'' with the lowest total weight~\cite{sahni1976p}.

In the optimization problems we study, we focus on the related problem of finding ``Hamiltonian paths'' instead. Similar to the TSP problem, we seek a path in an undirected weighted graph that visits each vertex exactly once but does not need to start and finish at the same vertex.

In Section~\ref{sec:gate_cancellation_implementation} we describe how the vertices, edges, and weights in the original Hamiltonian path problem map to our problem of interest: optimization for gate cancellation to get the minimal total number of gates.

The exact solution of TSP is is NP-hard and we cannot find a solution in polynomial time \cite{sahni1976p}. In  Section~\ref{sec:tsp_approx} we discuss TSP heuristics that can approximate the solution in polynomial time.

\section{Previous Work} \label{sec:previous_work}
Three major term-ordering techniques have been proposed by previous work: \textit{lexicographic} ordering, \textit{interleave} ordering, and \textit{magnitude} ordering. 

In particular, \textit{\textit{lexicographic}} ordering aims to provide a better solution for quantum gate cancellations. \textit{Interleave} ordering and \textit{magnitude} ordering, on the other hand, are more concerned with reducing Trotter errors, and the two share some similarities. In later sections, we will demonstrate that our grouping method can match the Trotter fidelities provided by \textit{magnitude} ordering and can achieve a greater gate cancellation factor than \textit{lexicographic} ordering does.

\subsection{Lexicographic Term Ordering}
Prior work has proposed the \textit{lexicographic} ordering to achieve better gate cancellation. The idea is to order the Pauli terms in alphabetic order. In particular, Hastings et al. \cite{hastings2014improving} propose lexicographic order at the level of fermionic operator strings (which is specific to molecular simulation use cases). Tranter et al. \cite{tranter2018comparison} broaden the lexicographic treatment to operate at the level of Pauli strings, thus broadening the applicability beyond molecular chemistry tasks.

For instance, assuming X $<$ Y $<$ Z $<$ I, the ordering for the following unordered string collection \{YYYYYXXZ, XXXXXYYX, YYYYYZII, XXXXXIXZ\} would be XXXXXYYX $<$ XXXXXIXZ $<$ YYYYYXXZ $<$ YYYYYZII. Then the corresponding quantum circuit will be constructed by concatenation of the Pauli terms in this order. We see that the higher-order characters are grouped next to each other. When applied to the time evolution quantum circuits, this ordering can provide many CNOT gates cancellations. In contrast, the original ordering would have no possible neighboring-term gate cancellation since all neighboring characters are different from each other. Using \textit{lexicographic} ordering, we can group the X's next to each other and Y's next to each other, resulting in many gate cancellation.

In general, \textit{lexicographic} ordering can provide a good level of gate cancellation since similar Pauli characters are grouped together, which  results in some similarities in the control and target qubits of CNOTs to cancel. Moreover, \textit{lexicographic} ordering can provide some improvement of Trotter fidelity, since similar terms imply similar physical properties.

\subsection{Interleave Term Ordering} \label{sec:interleave}
Hastings et al.~\cite{hastings2014improving} also propose the \textit{interleave} algorithm, which orders the Pauli terms in the following way, based on intuition from physics:
\begin{algorithm}[H]
\SetAlgoLined \SetKwData{FinalOrdering}{finalOrdering}
\KwIn{Molecular Hamiltonian in the Fermionic Representation}
\KwResult{Mapped Quantum Circuit in the Optimized Fermionic Ordering}
1. Execute all $H_{pp}$ and $H_{pqqp}$ terms

2. \For{$p, q$}{
          a. Execute $H_{pq}$ and all $H_{prrq}$ terms
    }
    
3. Execute all $H_{pqrs}$ terms

 \caption{\textit{Interleave} Ordering} \label{alg:interleave}
\end{algorithm}

The \textit{interleave} ordering improves Trotter fidelity because each step groups the terms with similar physical properties (this similarity is greater than what is provided by \textit{lexicographic} ordering) and all terms in lines 1 and 2a commute with each other, under the following condition \cite{hastings2014improving}:
\begin{equation}
    t_{pq} + \frac{1}{2} \sum V_{prrq} n_{r} = 0.
\end{equation}

\subsection{Magnitude Term Ordering}
Prior work has also has considered sorting the Pauli terms according to the magnitude of their coefficients \cite{hastings2014improving, tranter2018comparison}, namely, the absolute value of $h_{pq}$ and $h_{pqrs}$ defined in Eq.~\ref{eq:second_quantization}. 

The idea is that sorting the Pauli terms with similar coefficients magnitude will imply that Pauli terms with similar physical properties will be grouped together, thereby leading to fewer Trotter errors.

For instance, in \cite{hastings2014improving}, Hastings et al. observe that for the particular molecules they studied, the coefficient magnitude has the following ordering:
\begin{equation}
    h_{pp} > h_{pqqp} > h_{pq} > h_{pqqr} > h_{pqrs} .
\end{equation}
Therefore,  the \textit{magnitude} ordering likely can produce orderings similar to those of the \textit{interleave} ordering method.

\subsection{Deplete Grouping}
The \textit{depleteGroups} strategy was proposed by Tranter et al. \cite{tranter2019ordering}, which also partitions the Pauli terms into groups where every term commutes with every other term within the group. 
Once the Pauli terms are grouped into mutually commuting cliques, the final ordering is produced by iteratively selecting the highest magnitude term from each clique until all the groups have been exhausted, which is the opposite of the approach described in Section~\ref{sec:clique_ordering_abstract}.

\subsection{Random Ordering}
We also consider a \textit{random} ordering of the Pauli terms to serve as a baseline for comparison. Random term orderings were also used by Childs et. al.~\cite{childs2019faster} to prove stronger bounds on the size of the Trotter error.
\section{Multiterm Circuit Construction} \label{sec:circuit_construction}
In this section, we make several observations that justify the validity of using TSP gate cancellation (Section \ref{sec:gate_cancellation_implementation}) after the term-grouping preprocessing. This also provides some insight into the \textit{group-commutation} ordering strategy we discuss in more detail in Section \ref{sec:clique_ordering_abstract}. 
\subsection{Quantum Evolution Primitive}
Observation: For unitary evolution of the form $U(t) = U_{1}(t)U_{2}(t)$, suppose we have a quantum circuit A that efficiently simulates $U_{1}(t)$ and another quantum circuit B for $U_{2}(t)$. We can efficiently simulate $U(t)$ by concatenating circuit B and circuit A. For example, the evolution $U = e^{-i(aXX)t}e^{-i(bZZ)t}$ can be simulated by concatenating the $e^{-i(aXX)t}$ and $e^{-i(bZZ)t}$ circuits.

\subsection{Pairwise Commutation Circuit}
Theorem 1: For Hamiltonians of the form $H = H_{1} + H_{2}$, if $[H_{1}, H_{2}] = 0$, we have
\begin{equation}
    e^{-i(H_{1}+H_{2})t} = e^{-iH_{1}t}e^{-iH_{2}t}. \nonumber
\end{equation}

\begin{figure}[h!]
    \centering
    \includegraphics[scale=0.5]{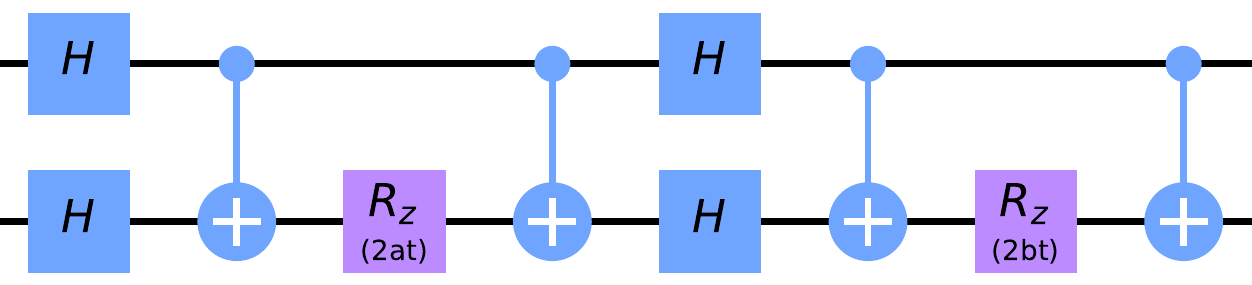}
    \caption{$U = e^{-i(aXX+bZZ)t}$ circuit}
    \label{fig:circ_XXZZ}
\end{figure}

Using Theorem 1, we can effectively construct the quantum circuit that simulates any two equal-sized Pauli terms that commute with each other. Figure~\ref{fig:circ_XXZZ} shows the circuit simulating the $U = e^{-i(aXX+bZZ)t}$ evolution.

\subsection{Group Commutation Circuit} \label{sec:group_circuit}

Theorem 2: For an arbitrary Hamiltonian of the form
\begin{equation}
    H=\sum_{n=1}^{M}H_{n}, \nonumber
\end{equation}
if 
\begin{equation}
    [H_{j}, H_{k}] = 0, \forall j, k \in [M], \nonumber
\end{equation}
then
\begin{equation}
    e^{-iHt} = e^{-iH_{1}t}e^{-iH_{2}t}...e^{-iH_{M}t}, \forall t.
\end{equation}

Corollary 2: For any $k, l \in \{1,...,n\}$, if $[H_{i}, H_{j}] = 0$ $\forall i, j \in \{1,...,n\}$, we have
\begin{multline}
     e^{-iH_{1}t}...e^{-iH_{k}t}...e^{-iH_{l}t}...e^{-iH_{n}t} = \\
    e^{-iH_{1}t}...e^{-iH_{l}t}...e^{-iH_{k}t}...e^{-iH_{n}t}.
\end{multline}

Theorem 2 is a powerful tool. If we can find a group of Pauli terms that mutually commute with all other terms in the group, we can construct the exact quantum circuit without any loss in Trotter fidelity. Therefore, it allows us to fully utilize the result we get from the classical  MIN-COMMUTING-PARTITION preprocessing (detailed explanation in the next section). 

\begin{figure}[t]
    \centering
    \includegraphics[scale=0.45]{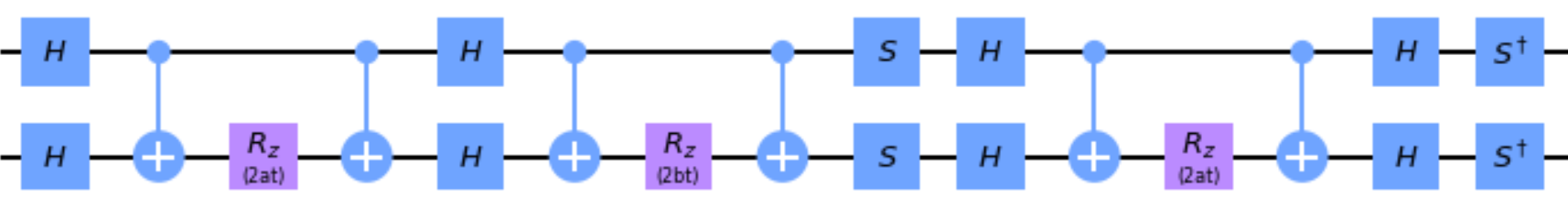}
    \caption{$U = e^{-i(aXX+bYY+cZZ)t}$ circuit}
    \label{fig:circ_XXYYZZ}
\end{figure}

As a simple example, we know that [XX, YY] = 0, [YY, ZZ] = 0, and [XX, ZZ] = 0. Therefore, we can construct the exact quantum circuit mapping of $U = e^{-i(aXX + bYY + cZZ)t}$ by concatenating them one after another, as shown in Fig.~\ref{fig:circ_XXYYZZ}. Corollary 2 then allows us to freely order the Pauli terms with a particular commuting partition without losing the overall Trotter fidelity. We can then formulate the problem of finding the optimal ordering inside each group as a TSP problem.

\subsection{Trotterization Circuit}
\label{sec:trotterization}
If the Pauli terms do not commute, we need to apply Trotterization/Trotter decomposition, justified by Lloyd \cite{Lloyd1996simulators}:
\begin{equation}
    e^{-iHt} \approx (e^{-iH_{1}t/n}e^{-iH_{2}t/n}...e^{-iH_{m}t/n})^{n}, \nonumber
\end{equation}
where $$H = \sum_{i=1}^{m} H_{i}.$$
More precisely \cite{whitfield2011simulation},
\begin{equation}\label{eq:first_order_trotter}
    e^{-iHt} = (e^{-iH_{1}\Delta t}e^{-iH_{2}\Delta t}...e^{-iH_{m}\Delta t})^{t/\Delta t} + O(t\Delta t).
\end{equation}

We denote $\frac{t}{\Delta t} = r$, also called the ``Trotter number.''

The approximation can be made tight by limiting the time step and Trotter number. The reason is that we have
\begin{equation}
    O(t\Delta t) = O(\frac{1}{r} t^{2}). \nonumber
\end{equation}
As $r \to \infty$ and $t \to 0$, the error in the approximation vanishes.

Higher-order approximations, such as the second-order Suzuki-Trotter approximation, also exist that further reduce the error. For demonstration purposes, we only implemented the first-order Trotter decomposition in our experimental simulation and theoretical examination. However, our techniques also apply for higher-order decompositions.

\subsubsection*{Trotterization Circuit Example}
Consider the following simple Hamiltonian for the deuteron \cite{dumitrescu2018cloud}.
\begin{equation}
    H = 5.907II + 0.2183ZI - 6.125IZ -2.143XX -2.143YY \nonumber
\end{equation}

\begin{figure}[h]
    \centering
    \includegraphics[width=0.45\textwidth]{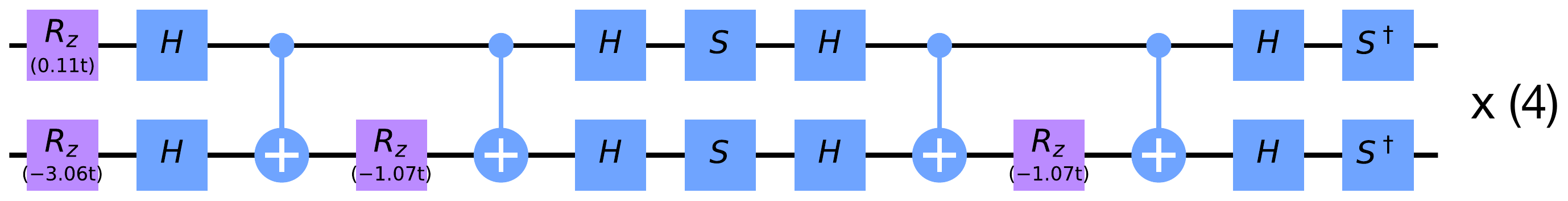}
    \caption{Quantum circuit simulating the quantum dynamics of deuteron, na\"ively partitioned. The entire circuit consists of the shown portion repeated 4 times.}
    \label{fig:deuteron_circuit}
\end{figure}

Na\"ively, we would apply Trotterization to all the Pauli terms.
Following Eq. \ref{eq:first_order_trotter}, we set the Trotter number to 4. We then have
\begin{equation}
    e^{-iHt} \approx (e^{\frac{-0.2183iZIt}{4}}e^{\frac{6.125iIZt}{4}}e^{\frac{2.143iXXt}{4}}e^{\frac{2.143iYYt}{4}})^{4} . \nonumber
\end{equation}
This corresponds to the quantum circuit in Fig.~\ref{fig:deuteron_circuit}.

\section{Group-commutation Term Ordering} 
\label{sec:clique_ordering_abstract}
In this section, we describe our \textit{group-commutation} ordering strategy that aims to achieve better total circuit fidelity and thus provide a desirable precondition for the TSP gate cancellation described in Section \ref{sec:gate_cancellation_implementation}.

\begin{figure}
    \centering
    \includegraphics[width=0.5\textwidth]{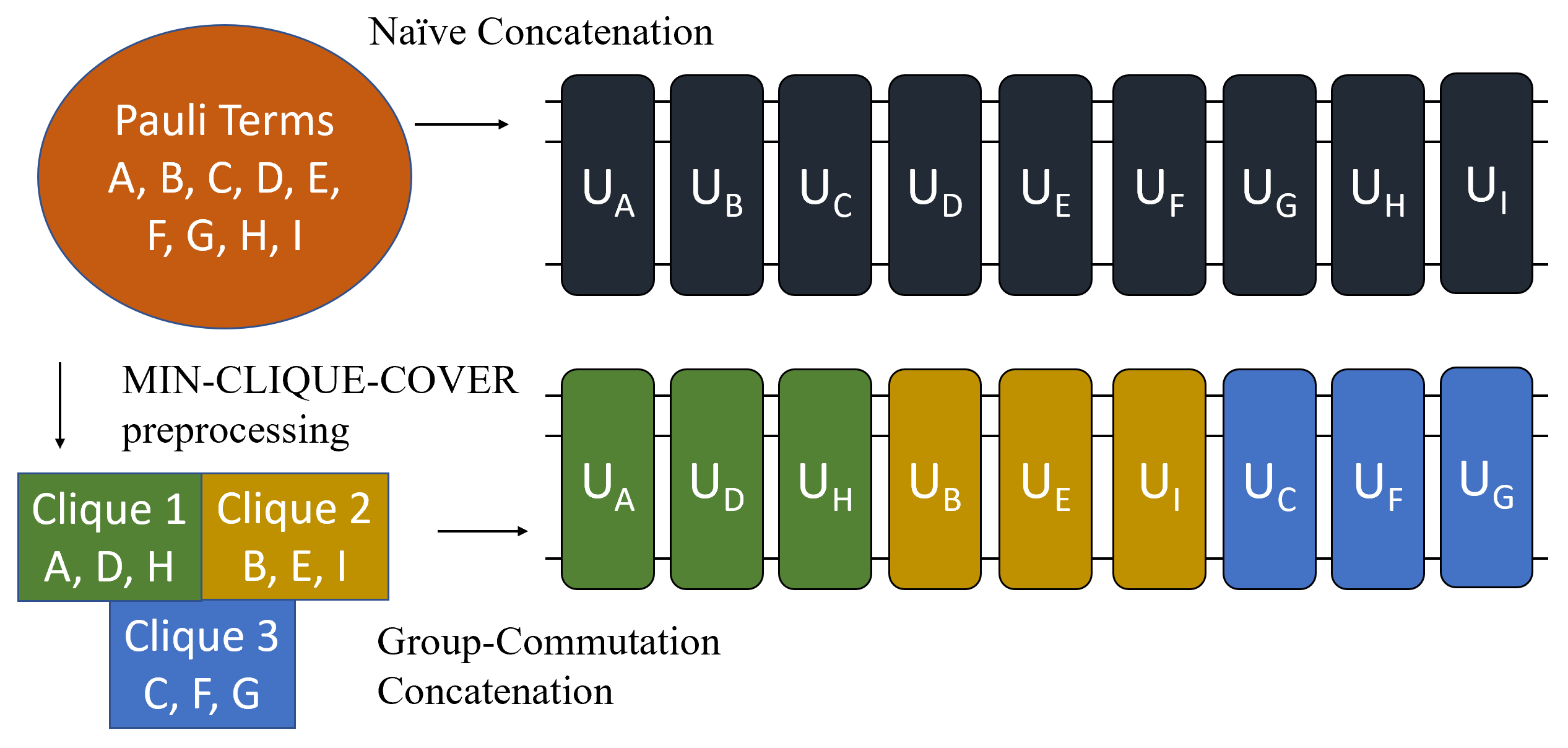}
    \caption{Group Commutation Ordering Example}
    \label{fig:circuit_concatenation}
\end{figure}

Intuitively, we know that if the commuting Pauli term pairs are grouped together in mutually commuting partitions, the Trotter errors can potentially be reduced. In other words, we are trying to first create the kind of concatenated sub-circuits that contain as much of the pairwise commuting Pauli terms as possible, shown in Section~\ref{sec:group_circuit}. Then we concatenate those sub-circuits one after another to construct the entire quantum circuit. An example of this concatenation method is shown in Fig~\ref{fig:circuit_concatenation}.

This intuition is explained more rigorously in Section \ref{sec:clique_ordering_theory}. In Section \ref{sec:clique_ordering_experimental}, we will experimentally demonstrate that the \textit{group-commutation} ordering is able to match the Trotter fidelity with that of \textit{magnitude} ordering, achieving much better Trotter fidelity than \textit{lexicographic} ordering does.

More importantly, not only can the \textit{group-commutation} method provide good fidelity, but it also has another benefit: after we arrange the Pauli terms into groups, we can freely arrange the term orderings in each group without affecting the fidelity (with the help of Corollary 2 in Section \ref{sec:circuit_construction}). Then we can consider the optimal ordering that maximizes gate cancellation inside each group.

\subsection{Problem Abstraction}
With this idea in mind abstract the problem as a graph problem. All the Pauli terms are represented as graph vertices, and an edge between a pair of vertices is introduced for commuting Pauli terms~\cite{gokhale2019minimizing}. 

Since the pairwise commutativity is not affected by each Pauli term's coefficient (while computing the term commutators, the coefficients are cancelled) and since the coefficients affect only the parameter of the Rz gates in the circuit mapping, we drop the coefficients of the Pauli terms in the initial grouping step (but will consider them later in the clique-clique ordering).

\begin{figure}[h]
    \centering
    \includegraphics[width=0.4\textwidth]{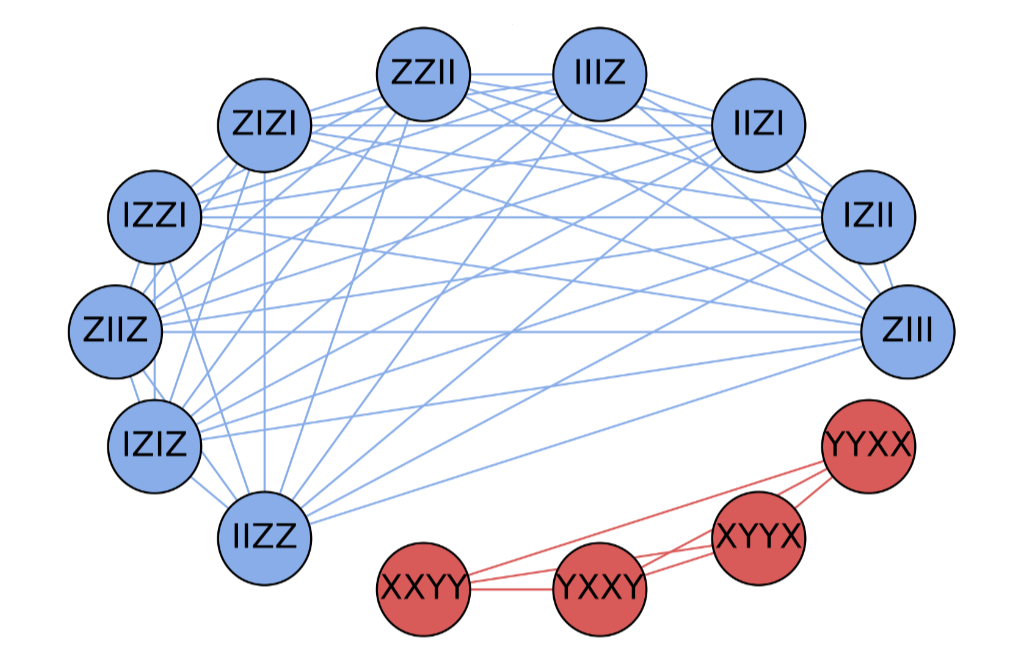}
    \caption{$H_{2}$ graph abstraction}
    \label{fig:H2_graph_abstraction}
\end{figure}

Consider the Hamiltonian for the $H_{2}$ molecule as an example. We obtain the following molecular Hamiltonian after applying the Jordan-Wigner transformation, and ignoring the IIII term that corresponds to a trivial circuit.
\begin{equation}
\begin{split}
    H = \\
    &0.0871IIIZ -0.0243 IIZI + 0.0871IZII \\
    &- 0.0243ZIII + 0.0785IIZZ + 0.135IZIZ \\
    &+ 0.0590XXYY + 0.0590YYYY + 0.0590XXXX \\
    &+ 0.0590YYXX + 0.138ZIIZ + 0.138IZZI \\
    & + 0.143ZIZI + 0.0785ZZII \nonumber
\end{split}
\end{equation}
This Hamiltonian is represented by the graph abstraction in \figref{fig:H2_graph_abstraction}. 
Note that the coefficients are not captured by the graph representation and they are not needed to determine commutativity relations. In this specific example, the 14 Pauli terms are separated into two fully connected subgraphs, or  cliques. Then, when mapping the problem to a quantum dynamics circuit, we can first concatenate all the IIIZ, IZIZ, ... terms in the blue colored clique, and then concatenate all the XXYY...YXXY terms in the red colored clique.

\subsection{Clique-Finding Algorithm}
We use the clique finding algorithm that we described in our earlier work~\cite{gokhale2019minimizing, gokhale2019n}. Given a Hamiltonian whose terms are a set of Pauli strings, we wish to partition the terms into commuting families where the number of partitions is minimized.

After representing the Hamiltonian as the graph described above, we then use the recursive Bron-Kerbosch \cite{bron1973algorithm} algorithm to find MIN-CLIQUE-COVERs on this graph. As presented in our previous work \cite{gokhale2019minimizing}, algorithms such as Boppana-Halldórsson \cite{boppana1992approximating} are able to provide polynomial time MIN-CLIQUE-COVER heuristics.

\subsection{Ordering the Cliques}\label{subsec:permutation_heuristic}

Finally, the last step in the \textit{grouping commutation ordering} is to select the ordering of cliques. Different orderings of the cliques can incur different Trotter errors by the same reasoning given above for the varying Pauli term orderings. Determining the optimal clique-clique ordering is intractable as it would require knowledge of the nested commutators between each of the cliques after the Baker–Campbell–Hausdorff expansion, i.e., an exponential amount of computation. Therefore, we use a polynomial-time heuristic to decide the clique-clique ordering that solely relies on the information from the approximated commutator (first-order approximated difference between $U_{approx}$ and $U_{exact}$ after the expansion). Note that, for two cliques $C_1 = \sum^{N_1}_{i=1} a_{i} A_{i}$ and $C_2 = \sum^{N_2}_{i=1} b_{i} B_{i}$, where $a_i, b_i \in \mathbb{R}$ and $A_i, B_i$ are Pauli terms, the commutator between them is
\begin{equation}
    [C_1, C_2] = C_1 C_2 - C_2 C_1 = \sum_{i=1}^{N_1} \sum_{j=1}^{N_2} a_i b_j (A_i B_j - B_j A_i).
    \label{eqn:clique-commutator}
\end{equation}

\begin{figure}[h!]
    \centering
    \includegraphics[width=0.95\columnwidth]{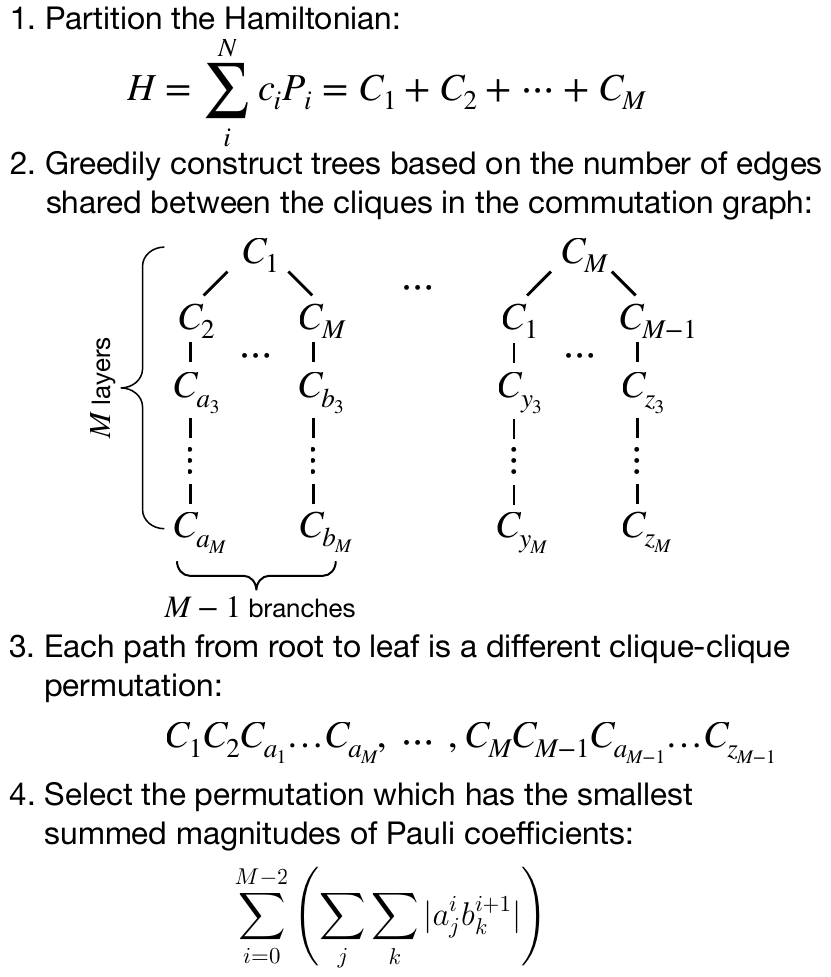}
    \caption{Polynomial time heuristic for selecting a clique-clique ordering. The commutation graph, which was constructed to partition the Hamiltonian into $M$ cliques (1), is utilized to greedily construct many trees (2) based on the number of edges shared between cliques. Each traversal of a tree from root to leaf produces a different clique-clique permutation (3). Finally, the permutation with the smallest commutator magnitude (Equation~\ref{eqn:clique-commutator}) is selected (4).}
    \label{fig:permutation_heuristic}
\end{figure}

Rather than computing the commutator between cliques exactly, the heuristic, shown in Figure~\ref{fig:permutation_heuristic}, exploits the information stored in the commutation graph that was used to group the Pauli terms into fully commuting cliques. Counting the number of edges between $C_1$ and $C_2$ in the commutator graph indicates the number of terms within the sum in Equation~\ref{eqn:clique-commutator} that evaluates to zero. The heuristic uses the number of inter-clique edges to greedily grow a tree (the nodes of the tree representing cliques) where each path through the tree corresponds to a potential clique-clique ordering. Intuitively, a permutation selected in this manner will produce commutators between the consecutive cliques that contain many zero terms. One can hope that this will reduce the overall magnitude of the commutator and therefore contribute very little to the overall Trotter error. 

Concretely, for each clique produced in the minimum clique cover, the heuristic constructs the tree described above with the current clique as the root. It then traverses each of these trees, producing a number of possible clique-clique permutations. Finally, for each permutation, it computes $\sum_{i=1}^{N_1} \sum_{j=1}^{N_2} |a_i b_j|$ over the non-zero terms and selects the permutation which produced the smallest magnitude, which one would expect to have the smallest contribution to the Trotter error. For a Hamiltonian which is partitioned into $M$ cliques, this heuristic has a runtime of $O(M^4)$ and produces $O(M^2)$ different permutations. We demonstrate the effectiveness of this heuristic in Section~\ref{sec:clique_ordering_experimental}. There can also be other types of heuristics for future explorations.
\section{Theoretical Analysis on 
Group-Commutation Ordering}
\label{sec:clique_ordering_theory}

In this section, we will justify the intuition that \textit{group commutation} ordering can potentially provide better Trotter fidelity than naive ordering for general Hamiltonians. In the next section, we will experimentally verify this intuition.

We analyze the effectiveness of our group commutation ordering strategy through the example of Hamiltonians with two commutation groups. The analysis can be generalized to more commutation groups by the reader. Consider a Hamiltonian $H= \sum_{i=1}^{k}\alpha_i H_i$, where $a_i$ are real numbers and $H_i$ are simple Hamiltonians that can be mapped to quantum circuits (or diagonalized) directly. Suppose $H$ can be divided into two commuting groups (cliques) $H^c_1 = \sum_{m=1}^{p} H_m$ and $H^c_2 = \sum_{m=p+1}^{k}H_m$, $i.e.$, $[H_m, H_n] = 0$ if $0<m, n \leq p$ or $p< m, n \leq k$. 

We compare the Trotter error of group commutation ordering and other orderings.\\

\textbf{With group commutation ordering}

It has been shown that \cite{knapp2005basic,dollard1979product, childs2019theory}, for $H$, the approximation error (in the additive form) of the Lie-Trotter formula is given by the variation-of-parameters formula,

\begin{align}
    \delta_{gc} &= e^{- i t(H_1^c)}e^{- i t(H_2^c)}-e^{-i t H} \nonumber\\
    &=\int_0^t d\tau e^{-i(t-\tau)H}[e^{-i\tau H^c_1}, H^c_2]e^{\tau H^c_2}.
    \label{eq:error}
\end{align}

Note that the error form also applies to general $H^g$ where  $H^g=H^g_1+H^g_2$. For simplicity, we denote the integral in Eq.~\ref{eq:error} as $I(H^g_1, H^g_2$) for general $H^g_1, H^g_2$. Thus, we can simply write the approximation error as $\delta_{gc} = I(H^c_{1}, H^c_2)$.

We are interested primarily in the operator norm ($i.e.$, spectral norm) of $||\delta_{gc}||= ||I(H^c_{1}, H^c_2)||$, which gives the worst-case analysis of the error.\\

\textbf{Without group commutation ordering}

 We can recursively apply the error formula in Eq.~\ref{eq:error} to the Lie-Trotter formula of arbitrary ordering for $H$.  Let $\pi$ be a permutation of the set $\{1,..,p\}$ that defines the ordering. First, we can approximate $e^{-i t H}$ by separating $H_{\pi(1)}$ from other terms:
\begin{align*}
    \delta_1 & = e^{-itH_{\pi(1)}}e^{-it\sum_{m=2}^kH_{\pi(m)}}- e^{-i t H}\\
    &=I(H_{\pi(1)}, \sum_{m=2}^kH_{\pi(m)}).
\end{align*}

Then we can recursively repeat the  process for the rest of the Hamiltonian and arrive at the following expression for the approximation error $\delta$ of the Lie-Trotter formula.

\begin{align*}
    \delta_{ngc} &= e^{-it H_{\pi(1)}}e^{-it H_{\pi(2)}}...e^{-it H_{\pi(2)}}-e^{-itH}\\
    &= I(H_{\pi(1)}, \sum_{m=2}^pH_{\pi(m)}) + e^{-it H_{\pi(1)}} I(H_{\pi(2)}, \sum_{m=3}^pH_{\pi(m)})\\
    & + e^{-it H_{\pi(1)}}e^{-it H_{\pi(2)}} I(H_{\pi(3)}, \sum_{m=4}^pH_{\pi(m)})+... \\
    &+ e^{-it H_{\pi(1)}}...e^{-it H_{\pi(j)}} I(H_{\pi(j+1)}, \sum_{m=j+2}^pH_{\pi(m)})+...
\end{align*}
Using the triangle inequality and the submultiplicativity of the operator norm, together with the fact that the operator norm of a unitary is 1, we have
\begin{align*}
    ||\delta_{ngc}|| &\approx 
   ||I(H_{\pi(1)}, \sum_{m=2}^pH_{\pi(m)}) +  I(H_{\pi(2)}, \sum_{m=3}^pH_{\pi(m)})\\
    & +  ... + I(H_{\pi(j+1)}, \sum_{m=j+2}^pH_{\pi(m)})+...|| .
\end{align*}

Also we know that $H_{\pi(k)}$ is either in $H^c_1$ or $H^c_2$. Thus, we have
\begin{align*}
    ||\delta_{ngc}||& \approx ||\sum_{\pi(j) \in [1,p]} I(H_{\pi(j)}, H^c_2)||+\\
    &||\sum_{\pi(j) \in [p+1,k]} I(H_{\pi(j)}, H^c_1)||.
\end{align*}

Although we have no proof that $||\delta_{gc}|| < ||\delta_{ngc} ||$ (because we do not have  information about the full commutation relation and magnitude information in $H$), we can make several observations why group commutation ordering is advantageous. First, $||\delta_{gc}||$ in general has a much lower upper bound than does the first term in $||\delta_{ngc}||$. In fact, the upper bound of $||\delta_{gc}||$  does not scale with the number of terms $p$ and that of $||I(H_{\pi(1)}, \sum_{m=2}^pH_{\pi(m)}) ||$ is of $O(p)$. Second, $||\delta_{gc}||$ does not include the second term in $||\delta_{ngc}||$. Thus, there is strong evidence that \textit{group commutation} ordering has an advantage over naive ordering in terms of eliminating Trotter errors.

In the next section, we will support the above theoretical intuition by implementing real molecular Hamiltonian and test their Trotter fidelities versus other ordering methods such as \textit{lexicographic} and \textit{magnitude} orderings.

\section{Experimental Demonstration of Group-Commutation Ordering} \label{sec:clique_ordering_experimental}

In this section, we experimentally demonstrate the fact that \textit{group-commutation} ordering can provide total circuit fidelities similar to the fidelities provided by \textit{magnitude} ordering, thus preparing for the in-clique TSP gate cancellation described in Section \ref{sec:gate_cancellation_implementation}. We also show that different permutations of the cliques will lead to different total circuit fidelities. Lastly, we show the effect of our clique-clique ordering strategies discribed in section~\ref{subsec:permutation_heuristic}.

\subsection{Methodology}\label{subsec:fidelity_method}
To show that the group commutation circuit can provide better fidelity, we construct an example DQS circuit to simulate a simple 2-qubit Hamiltonian with its terms sorted according to group commutation,
\begin{equation}
    H_{C} = IZ + ZI + ZZ + XX + YY, \nonumber
\end{equation}
and with terms sorted in some random order,
\begin{equation}
    H_{R} = XX + ZI + YY + IZ + ZZ .\nonumber
\end{equation}
The process fidelity metric~\cite{salathe2015digital, harper2020efficient, qiskit2020process, gilchrist2005distance} measures how closely the DQS circuit's unitary matrix, $U_{approx}$, matches the exact unitary evolution matrix, $U_{exact} = e^{-i H t}$,

\begin{equation}\label{eq:fidelity}
    \mathcal{F} = \frac{\lvert \Tr(U_{exact} U_{approx}^{\dag})\rvert}{\dim(U_{exact})}
\end{equation}

\begin{figure}[h!]
    \centering
    \includegraphics[scale=0.4]{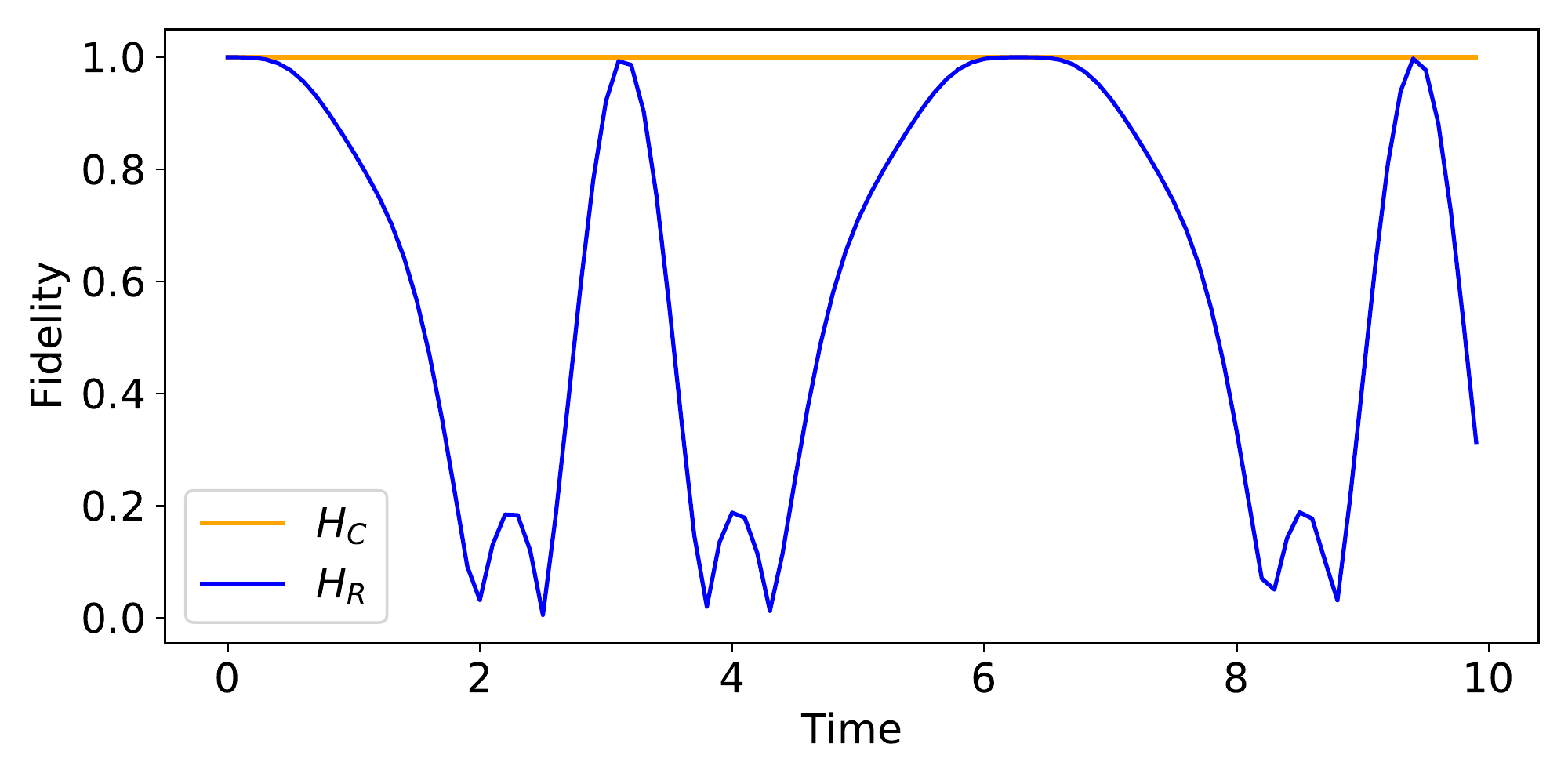}
    \caption{The same Hamiltonian is simulated using two different term orderings $H_C$ and $H_R$.}
    \label{fig:H2_without_Co}
\end{figure}

The result is plotted in Fig. \ref{fig:H2_without_Co}, with different time iteration t values. The orange curve (group ordering) is a straight line that achieves perfect fidelity matching the exact unitary evolution. On the other hand, the random ordering fidelity represented by the blue curve oscillates between 0 and 1 due to Trotter errors.

The group ordering places the two all-pairwise commuting groups of Pauli terms $\{XX, YY, ZZ\}$ and $\{ZI, IZ\}$ together. We also observe that\\ $[(XX + YY + ZZ), (ZI + IZ)] = 0$. Therefore, applying Theorem 2 twice, we can explain why the resulting circuit achieves perfect fidelity.

\subsection{Benchmarking of Molecular Hamiltonian}

For a typical Hamiltonian, the groups of Pauli terms do not commute with each other with coefficients added. Thus, in the general case we are unlikely to achieve the perfect fidelity as depicted in Fig. \ref{fig:H2_without_Co}. 

We first use the $\text{LiH}$ molecule to demonstrate the effectiveness of term grouping and the importance of clique-clique ordering. We note that the results of other molecules are also behaving similarly. We first partitioned the 26 terms into $n$ maximally commuting cliques and then enumerated all $n!$ permutations of the cliques. For each permutation we constructed a DQS circuit to simulate the Hamiltonian evolution and computed this circuit's fidelity according to \meqref{eq:fidelity}, with time iteration step t = 0.01 and Trotter number r = 10.

\begin{figure}[h]
    \centering
    \includegraphics[scale=0.34]{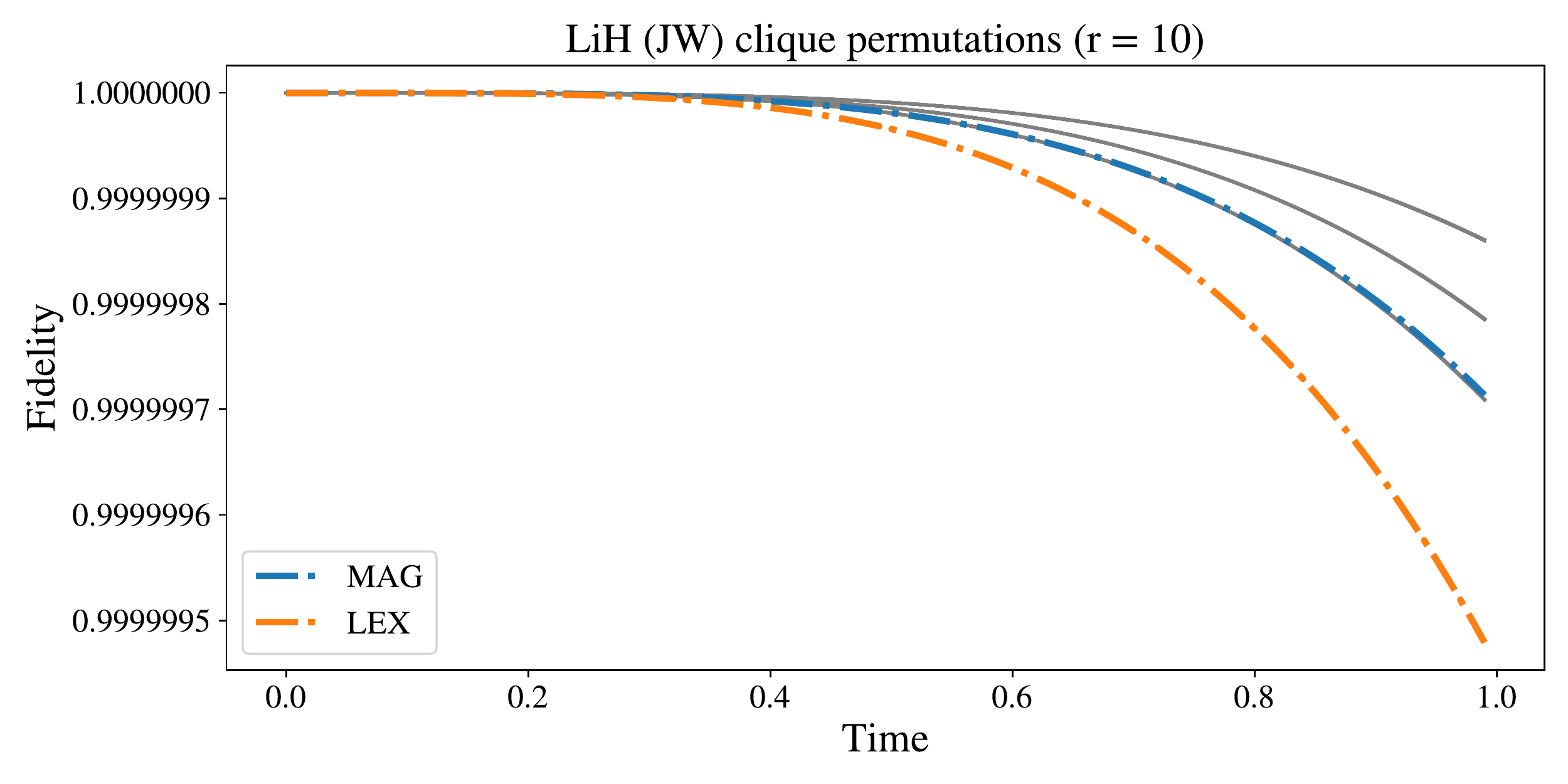}
    \caption{LiH Hamiltonian partitioned into 4 commuting cliques. The resulting 24 permutations for the Hamiltonian are shown as gray solid lines while the \textit{magnitude} and \textit{lexicographic} orderings are dotted. The ``Time'' in the x-axis is a unitless parameter.}
    \label{fig:lih_timeseries}
\end{figure}

\begin{figure}[h]
    \centering
    \includegraphics[scale=0.34]{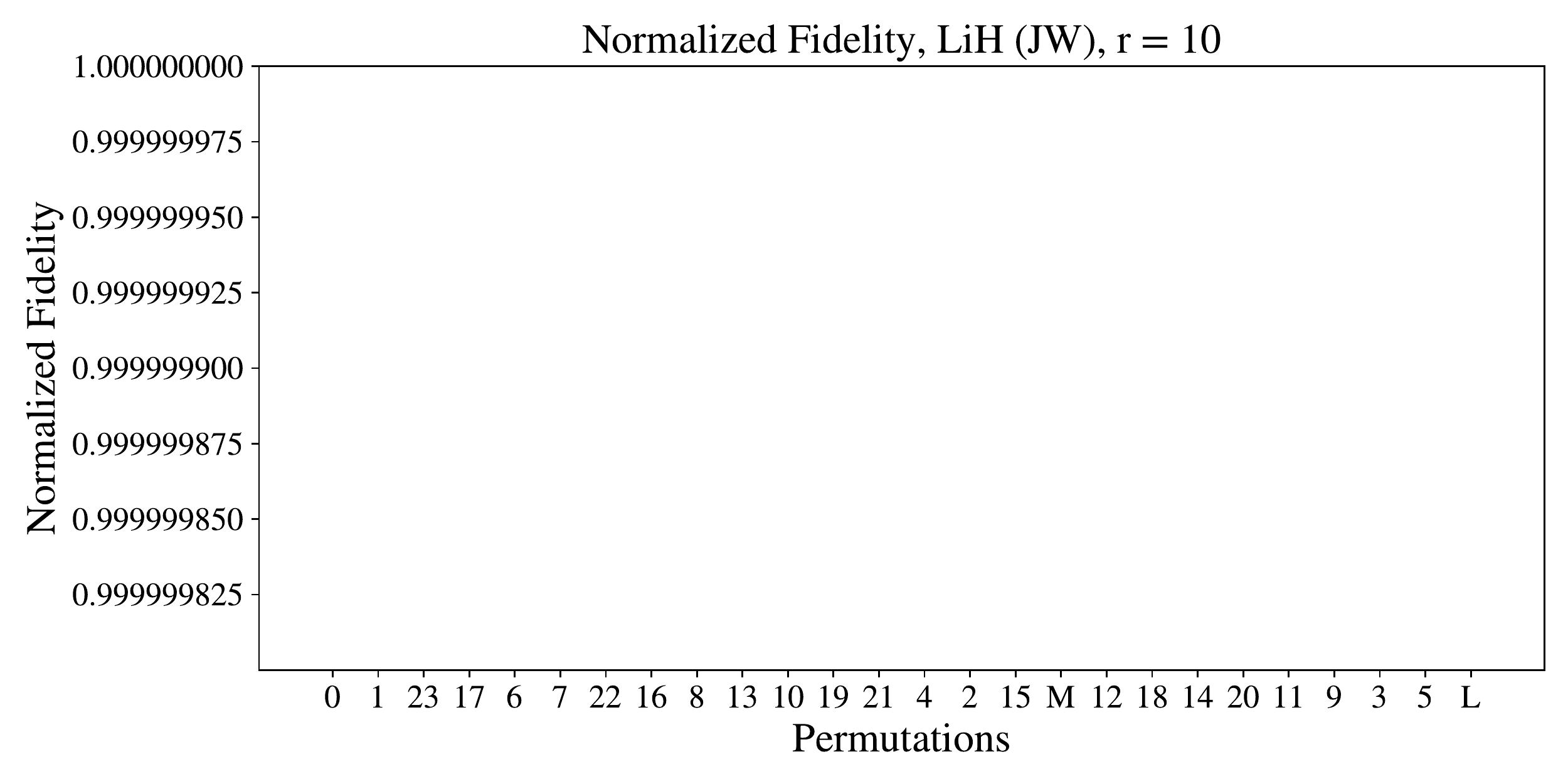}
    \caption{Normalized fidelity as defined by Eq. \ref{eq:fidelity_integral} for various permutations of the Hamiltonian. Fidelities of random permutation orderings are shown in blue and Fidelities of the \textit{magnitude} and \textit{lexicographic} orderings are shown in orange. }
    \label{fig:lih_fidelity}
\end{figure}

Figures~\ref{fig:lih_timeseries} show the results of the process fidelity time series for the LiH molecule whose 26 Pauli terms that can be grouped into 4 commuting cliques. \figref{fig:lih_fidelity} shows the fidelities of different permutations (blue) as well as the fidelities of \textit{magnitude} and \textit{lexicographic} orderings (orange). The normalized fidelity values in \figref{fig:lih_fidelity} were computed (using the fidelities $\mathcal{F}(t)$ in \figref{fig:lih_timeseries}) according to
\begin{equation}\label{eq:fidelity_integral}
    \frac{\int_0^{t'}{\mathcal{F}(t) dt}}{\int_0^{t'}{dt}}.
\end{equation} We use these benchmark results to show that the best clique-clique permutations are able to achieve fidelities on a par with the \textit{magnitude} ordering.

\subsection{Effects of Clique-Clique Ordering Heuristics}
\label{subsubsec:clique_clique_ordering_exp}

\begin{figure}[h!]
  \centering
  \includegraphics[width=0.9\columnwidth]{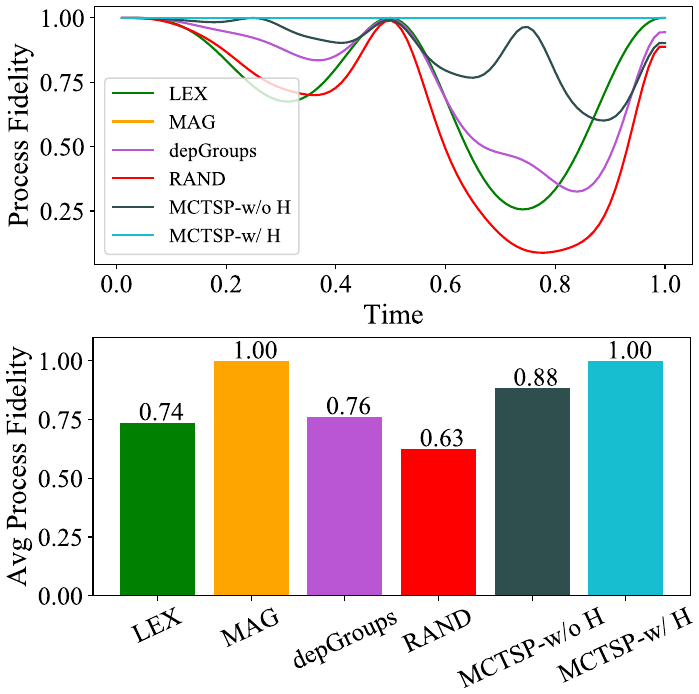}
  \caption{Process fidelity for CO\textsubscript{2} over a range of values for the unitless time parameter. Two \textit{max-commute-tsp} orderings are shown, one which utilizes the clique-clique permutation heuristic from Section~\ref{fig:permutation_heuristic} and one which does not. Performance is greatly improved when the heuristic is utilized (the MCTSP-w/H and MAG results overlap one another).}
  \label{fig:co2_timeseries}
\end{figure}

Here we show the additional effects of our clique-clique ordering heuristics on Trotter fidelities, using the process fidelity time series of DQS for the CO\textsubscript{2} molecule Figure~\ref{fig:co2_timeseries}. We note that the heuristic has a similar effect on all of the benchmark Hamiltonians (shown in section~\ref{sec:combined_benchmarking}. Without the heuristic, \textit{max-commute-tsp}'s process fidelity varies over time, but with the heuristic, its curve overlaps that of the \textit{magnitude} ordering at a constant process fidelity of 1.0. As the number of Pauli terms in the Hamiltonians grows, so to will the number of cliques, and therefore a good permutation heuristic will be required to select a single clique-clique ordering from factorially many choices.

\subsection{Result Discussion}
Although both the \textit{group-commutation} and \textit{magnitude} orderings achieve similar fidelities, the number of possible gate cancellations differs greatly between the two ordering schemes. The \textit{group-commutation} ordering is especially well suited for gate optimizations because of the freedom it provides for term orderings within each clique. In \secref{sec:gate_cancellation_implementation} we describe how we take advantage of this fact to order terms within cliques, which is optimal for gate cancellations. On the other hand, a \textit{magnitude} ordering is determined by term coefficients that does not carry any information on the gate cancellations between terms, and therefore not expected to have good gate cancellation performances.

For other physical systems (i.e., Hamiltonians that describe solid-state structures, high energy physics, etc.), the fidelity comparison would still require investigation. For instance, if we have a Hamiltonian that has Pauli terms all with similar magnitude coefficients, \textit{group-commutation} ordering might offer a much better total circuit fidelity. Moreover, the \textit{magnitude} and \textit{interleave} ordering are based on the Hamiltonians that evolve only the single and double electron integrals defined in Eq. \ref{eq:second_quantization}. Other Hamiltonians of physical systems with different quantization structures are likely to have different results.

\section{Pairwise Gate Cancellation Tricks and Restrictions} \label{sec:pairwise_gate_cancellation}
In this section we first provide some of the important gate cancellation techniques and restrictions. Then we show a useful ``\textit{star + ancilla}'' trick that fully utilizes the gate cancellation techniques and at the same time is able to bypass the restrictions. The next section (Section \ref{sec:gate_cancellation_implementation} will describes how we actually formulate these tricks and restrictions, together with the \textit{star + ancilla} trick, as a TSP problem, after we group the Pauli terms into cliques using the \textit{group-commutation} ordering described in Section \ref{sec:clique_ordering_abstract}.

\subsection{Gate Cancellation Techniques}
We observe that many cancellations of the gates specified in Section \ref{sec:gate_basics} are possible in the quantum dynamics circuit.

\subsubsection{Cancellation of Single-Qubit Gates}
First, we observe that two Hadamard gates acting on the same qubit consecutively can cancel each other since $HH = I$. This will happen when two Pauli X characters or an X and a Y character are next to each other.

We can also cancel the gate sequence $\{S,H,H,S^{\dagger}\}$ when two Pauli Y characters are next to each other, since $SHHS^{\dagger} = SS^{\dagger} = I$.

\subsubsection{Cancellation of Multiqubit Gates}
Next, we observe that two ideal CNOT gates acting on the same control and target qubits can cancel each other. We can prove this cancellation by multiplying two CNOT matrices.

\begin{equation}
\begin{pmatrix}
1 & 0 & 0 & 0 \\
0 & 1 & 0 & 0 \\
0 & 0 & 0 & 1 \\
0 & 0 & 1 & 0
\end{pmatrix}
*
\begin{pmatrix}
1 & 0 & 0 & 0 \\
0 & 1 & 0 & 0 \\
0 & 0 & 0 & 1 \\
0 & 0 & 1 & 0
\end{pmatrix}
=
\begin{pmatrix}
1 & 0 & 0 & 0 \\
0 & 1 & 0 & 0 \\
0 & 0 & 1 & 0 \\
0 & 0 & 0 & 1
\end{pmatrix}
= I \nonumber
\end{equation}

In addition, when other gates are in between the CNOT gate pairs, we can do a ``swap \& cancel'' if the CNOT gate matrix commutes with the other gate matrices. The commutation relation does not always hold, however, as we explain next.
\subsection{Gate Cancellation Restrictions}
Certain gate orderings can prevent the gate cancellations describe above. We identify several kinds of these restrictions below.
\subsubsection{Restriction I}
Two ways of implementing the CNOT entangler exist. The first is to put the target qubits of CNOT gates on all the qubits, as used in the quantum dynamics circuit construction in Section \ref{sec:background} and \ref{sec:circuit_construction}. We denote this the ``\textit{ladder}'' implementation. The second way is to put the target qubits of CNOT gates on one qubit. We denote this the ``\textit{star}'' implementation. The two implementations will yield identical outputs.

\begin{figure}
    \centering
    \includegraphics[width=0.3\textwidth]{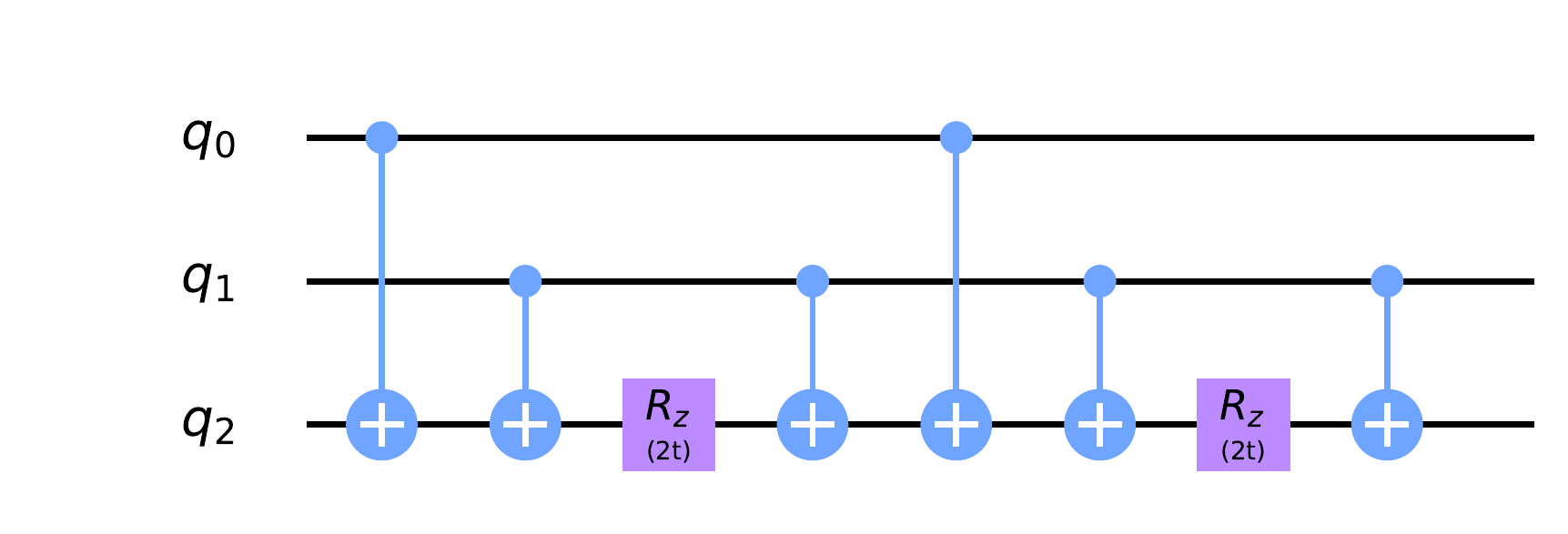}
    \caption{\textit{Star} implementation original circuit}
    \label{fig:Center_Original}
\end{figure}

\begin{figure}
    \centering
    \includegraphics[width=0.3\textwidth]{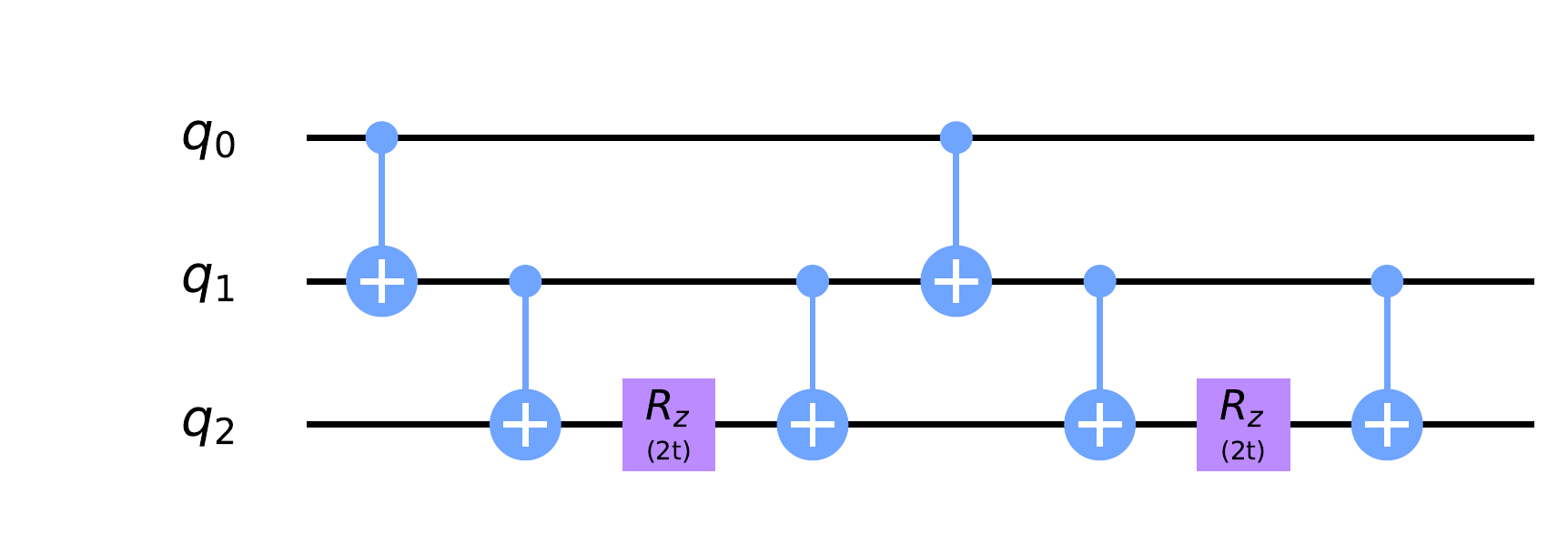}
    \caption{\textit{Ladder} implementation original circuit}
    \label{fig:Ladder_Original}
\end{figure}

Consider the following Hamiltonian $H = ZZZ + ZZI$ as an example. Its two implementations are shown in Fig.~\ref{fig:Center_Original} and Fig.~\ref{fig:Ladder_Original}

\begin{figure}
    \centering
    \includegraphics[width=0.25\textwidth]{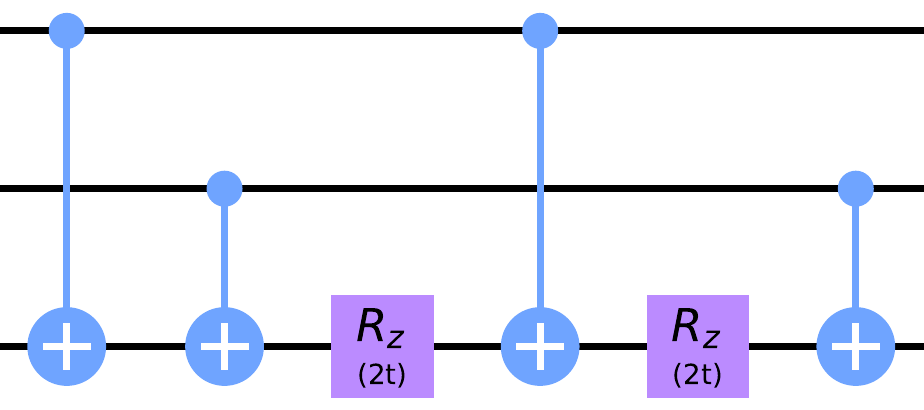}
    \caption{\textit{Star} implementation optimized circuit}
    \label{fig:Center_Optimized}
\end{figure}

We can cancel the third and fifth CNOT gates in the \textit{star} implementation (Fig.~\ref{fig:Center_Optimized}) but not in the \textit{ladder} implementation. Canceling the third and fifth CNOT gates in the \textit{ladder} implementation would result in a different quantum circuit output.

\subsubsection{Restriction II}
Sometimes when Hadamard gates exist between the CNOT sequences (resulting from X or Y characters in the Pauli terms), we cannot cancel certain pairs of CNOT gates. The reason is that H and the NOT part of CNOT do not commute with each other.
\begin{equation}
[H, X] = HX - XH
= \sqrt{2}
\begin{pmatrix}
0 & 1 \\
-1 & 0
\end{pmatrix}
\neq 0 \nonumber
\end{equation}
The fact that H and X do not commute implies that we cannot simply swap the H and CNOT gates. Therefore, we cannot perform the ``swap and cancel'' technique in the previous example for the CNOT pairs that have the same control and target qubits. 

\begin{figure}
    \centering
    \includegraphics[width=0.45\textwidth]{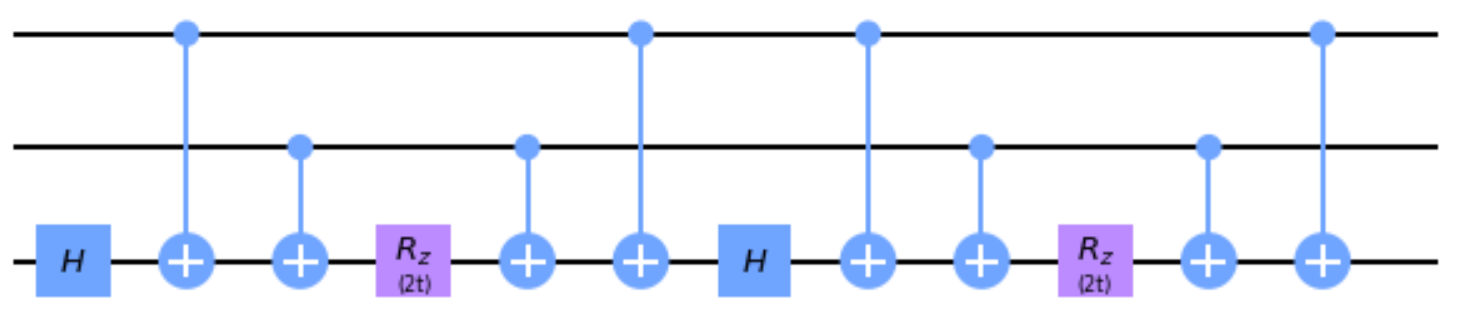}
    \caption{Hadamard restriction}
    \label{fig:H_restriction}
\end{figure}

Consider the following example of the Hamiltonian $H = XZZ + ZZZ$ (Fig.~\ref{fig:H_restriction}).
We cannot cancel the third \& sixth CNOT and the fourth \& fifth CNOT pairs because there is a Hadamard gate in between.

Similarly, if the Hadamard gate is operating on the same qubit as one of CNOT's control qubits, we also cannot swap the Hadamard with the CNOT, because
\begin{equation}
    [H\otimes I, CNOT] \neq 0 . \nonumber
\end{equation}
\subsubsection{Restriction III}
In some cases CNOT target qubits are not the same for different Pauli terms. 

Consider the Hamiltonian $H = ZZZZIIII + IIIIZZZZ$ as an example. The target qubit of the $ZZZZIIII$ term can never overlap with the $IIIIZZZZ$ term, and hence no possible CNOT gate cancellation can occur.

\subsection{Star + Ancilla Trick for CNOT Cancellation} \label{subsec:cnot_cancellation_trick}
\begin{figure}[h]
    \centering
    \includegraphics[width=0.45\textwidth]{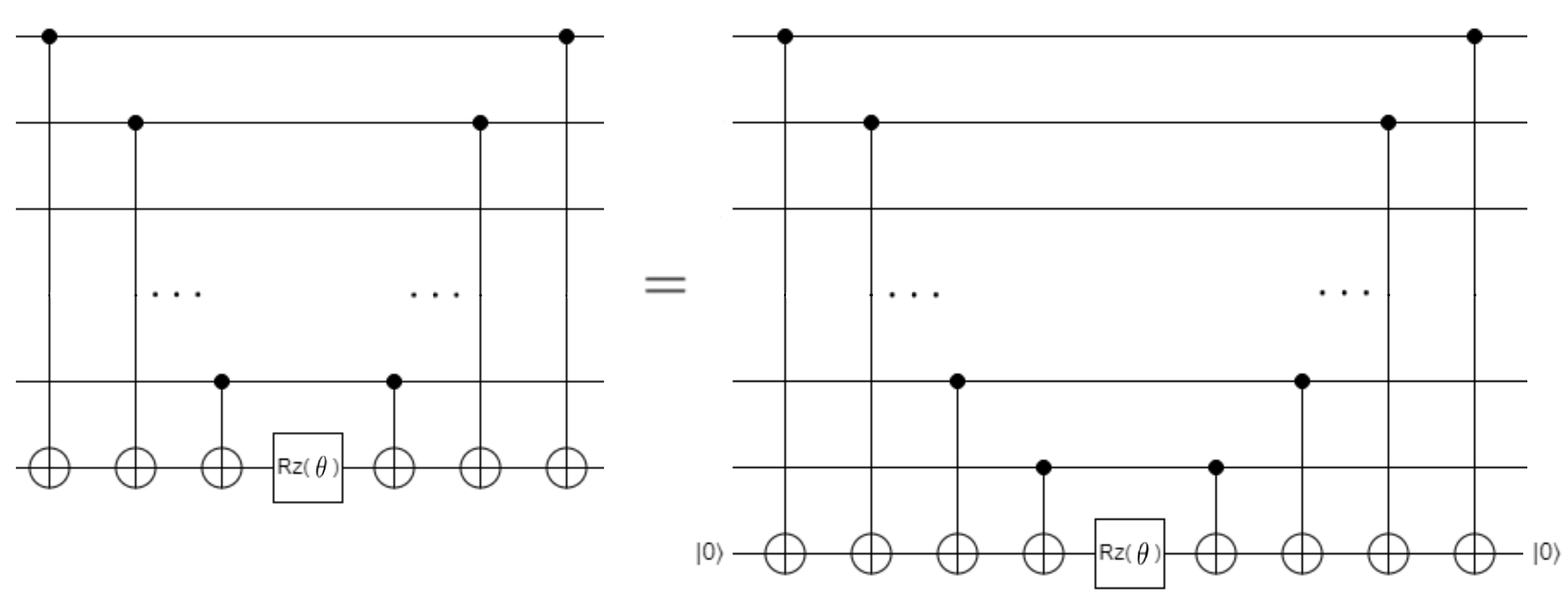}
    \caption{\textit{Star + Ancilla} Trick}
    \label{fig:trick_demo}
\end{figure}

One interesting trick presented in \cite{nielsen2002quantum} suggested that we can move the target qubit of the CNOT entanglers to an additional zero-initialized qubit by adding two CNOT gates (Fig.~\ref{fig:trick_demo}). This trick allows us to bypass all the restrictions described above.

\subsubsection{Bypassing the Restrictions}

\begin{figure}
    \centering
    \includegraphics[width=0.45\textwidth]{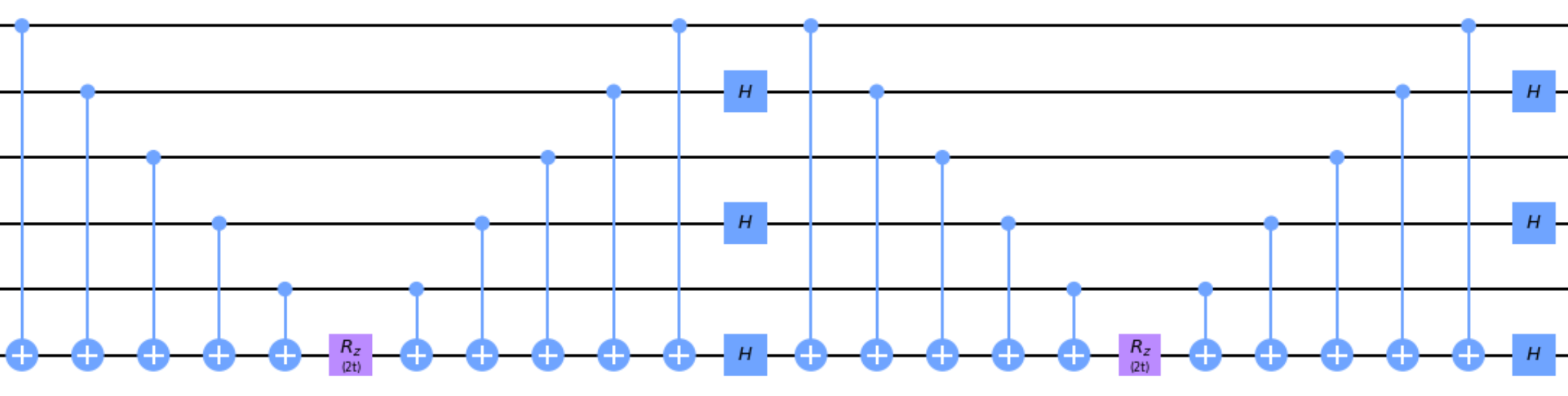}
    \caption{ZZZZZZ \& ZXZXZX original \textit{star} implementation}
    \label{fig:trick_star_original}
\end{figure}

\begin{figure}
    \centering
    \includegraphics[width=0.45\textwidth]{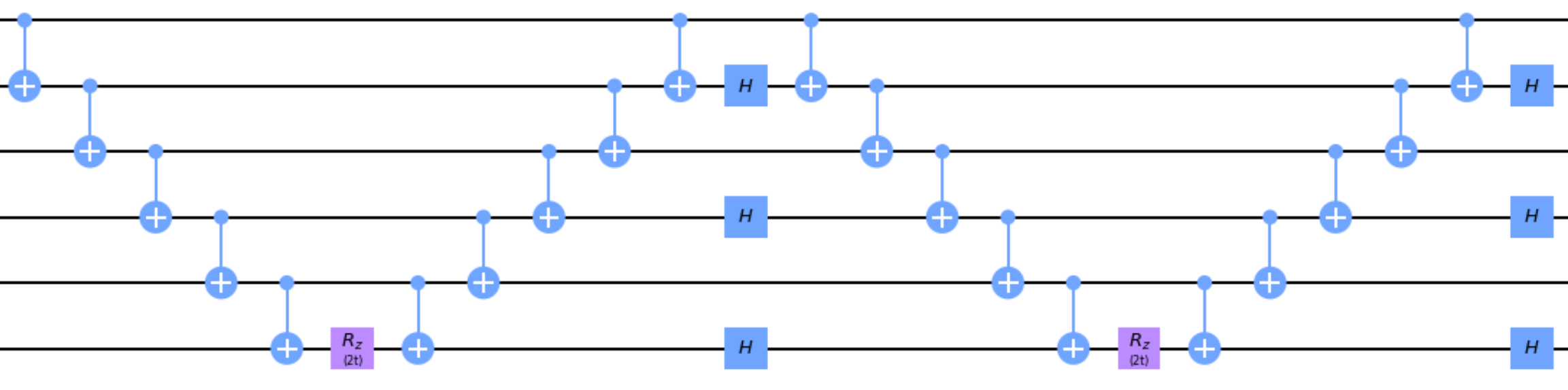}
    \caption{ZZZZZZ \& ZXZXZX original \textit{ladder} implementation}
    \label{fig:trick_ladder_original}
\end{figure}

Consider the example of the two Pauli terms ZZZZZZ and ZXZXZX. The original \textit{star} implementation (Fig.~\ref{fig:trick_star_original}) is unable to have any CNOT cancellation because there is one Hadamard acting on the sixth/last qubit that blocks any swaps and cancellation of the CNOTs.
The CNOT pairs in the original \textit{ladder} implementation (Fig.~\ref{fig:trick_ladder_original}) are also unable to cancel with each other, since there are always Hadamard gates standing on either the target or the control qubit of the CNOTs.

\begin{figure}
    \centering
    \includegraphics[width=0.48\textwidth]{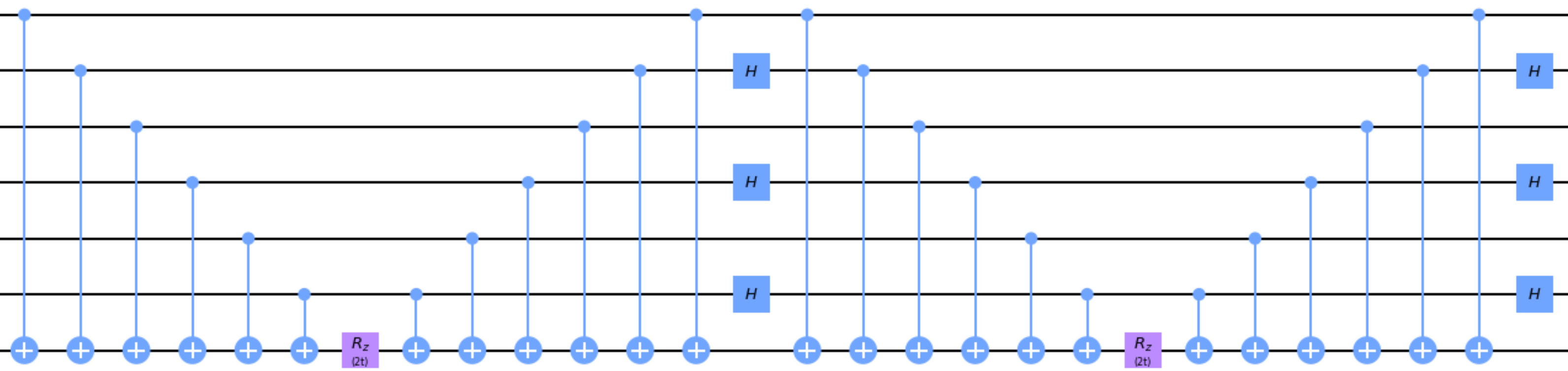}
    \caption{ZZZZZZ \& ZXZXZX modified \textit{star} implementation}
    \label{fig:trick_star_new}
\end{figure}

\begin{figure}
    \centering
    \includegraphics[width=0.45\textwidth]{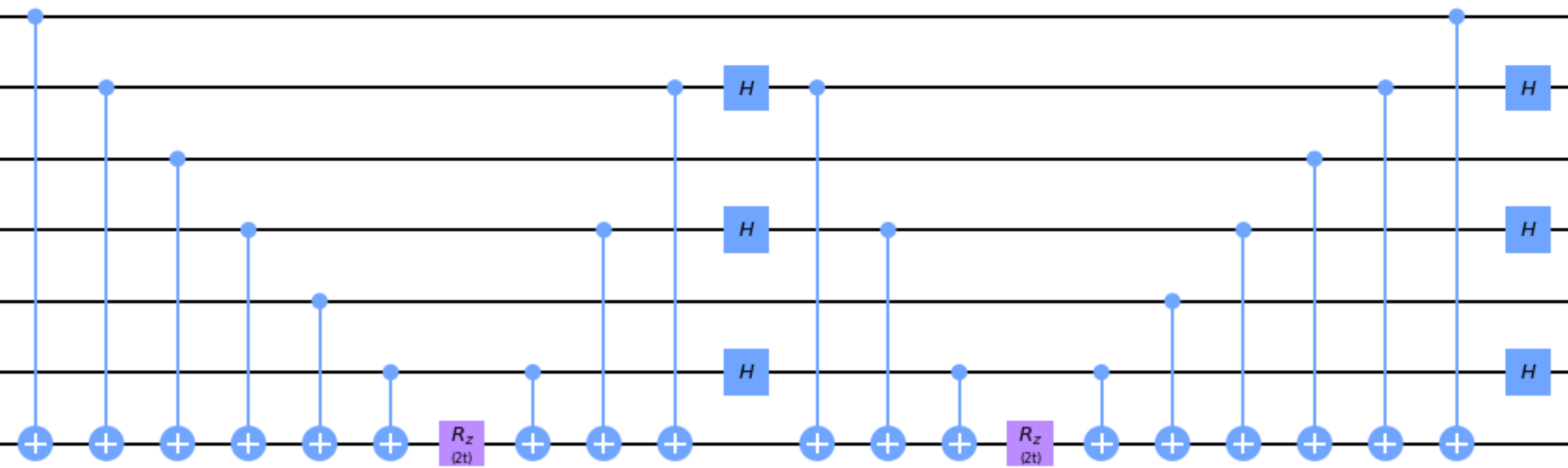}
    \caption{ZZZZZZ \& ZXZXZX modified \textit{star} implementation with gate cancellation}
    \label{fig:trick_star_new_cancelled}
\end{figure}

With the help of the trick, we can first modify the \textit{star} implementation to the circuit shown as Fig.~\ref{fig:trick_star_new}. 
Then the middle six CNOT gates are canceled (Fig.~\ref{fig:trick_star_new_cancelled}), thus solving Restriction II.
Having the trick also allows us to move all target qubits to the ancilla qubit and thus allows us to do cancellation of CNOT gates. Thus, Restriction III is also solved. Since the trick is a \textit{star} implementation by definition, Restriction I is inherently bypassed.

Although we add four more CNOT gates and one more ancilla qubit in Fig.~\ref{fig:trick_star_new}, this eventually leads to six CNOT gate cancellations. In the more general cases where the Pauli terms are very long, this trick can provide greater cancellation numbers, since the additional cost in Fig.~\ref{fig:trick_star_new} is constant (2 additional CNOT gates for each Pauli term and only one ancilla qubit that works for the entire circuit)
\section{Gate Cancellation via TSP} \label{sec:gate_cancellation_implementation}

In this section, we describe a strategy for maximizing gate cancellation in a quantum dynamics circuit, in order to reduce its hardware implementation cost. We focus primarily on the cancellation of two-qubit CNOT gates, which are the dominant source of errors and latency in the prevalent quantum platforms. Typically, CNOT gates have at least 10x lower fidelity and 2--5x longer duration than single qubit gates \cite{linke2017experimental, debnath2016demonstration, gambetta2019cramming}. We also show that TSP provides better gate cancellation results than \textit{lexicographic} ordering under the same circuit Trotter fidelities. Since \textit{magnitude} ordering does not concern the gate cancellation regime, TSP and \textit{lexicographic} perform much better than \textit{magnitude} ordering.

The task we perform is ordering a set of $k$ Pauli strings, $\{P_1, P_2, ..., P_k\}$, to maximize the CNOT gate cancellation in the corresponding quantum dynamics circuit. To simplify our notation in this discussion, we have ignored the coefficients of the Pauli strings, because they do not affect the CNOT cancellation properties. We assume that any permutation of the Pauli strings, $\pi: [1, 2, ..., k] \to [1, 2, ..., k]$, is valid. This is indeed the case when starting from the \textit{group-commuting} ordering method described in Section~\ref{sec:clique_ordering_abstract}, since the resulting Trotter error is zero, invariant of term ordering (Corollary 2). Alternatively, one may decide that the gate cancellation gains outweigh the Trotter error variances. In that case each permutation could be treated as valid.

Each of the $k!$ permutations corresponds to a quantum circuit that implements
\begin{equation} \prod_{j=1}^k \exp(-i P_{\pi(j)} t). \nonumber \end{equation}
We focus on the ``\textit{star + ancilla}'' circuit architecture, described in Section~\ref{subsec:cnot_cancellation_trick} and illustrated in Figure~\ref{fig:trick_demo}), because this circuit architecture simplifies the gate cancellation logic. Each non-$I$ Pauli character originates two identical CNOTs, controlled on the non-$I$ index and targeting the ancilla qubit. These two CNOTs are symmetric around the middle $R_z$ on the ancilla. Without any gate cancellation, we know that the number of CNOT gates required for $\exp(-i P_{\pi(j)} t)$ is
\begin{equation} 2 \sum_{j=1}^k |P_j|_\text{Ham} = 2 \sum_{j=1}^k \sum_{i=1}^N \mathbbm{1}_{P_j[i]\neq I}, \label{eq:cnot_upper_bound} \end{equation}
where $N$ denotes the width of the Pauli strings and Ham refers to Hamming weight.

However, a good permutation $\pi$ can substantially reduce the CNOT count against Eq.~\ref{eq:cnot_upper_bound}'s upper bound because the CNOT gates between neighboring $\exp(-i P_{\pi(j)} t)$ subcircuits can cancel each other. Concretely, let us consider two consecutive Pauli strings, $[P_1, P_2]$. We focus on the ``transition zone'' from $\exp{(-i P_1 t)}$ to $\exp{(-i P_2 t)}$. This spans the CNOT gates between the $R_z$ gates for $P_1$ and for $P_2$, and this span is where we can orchestrate gate cancellation. From the previous analysis, prior to gate cancellation, the number of CNOT gates required for the transition zone is
\begin{equation} |P_1|_{\text{Ham}} + |P_2|_{\text{Ham}}. \label{eq:transition_CNOTs} \end{equation}

For two neighboring CNOTs to cancel each other, they must be identical, in other words have the same control and target. Moreover, as described in Section~\ref{sec:pairwise_gate_cancellation}, the single-qubit gates on the controls must also cancel out---this requires the Pauli characters of the controls to be identical. There are no other restrictions, since the shared-target CNOT blocks are order insensitive. Intuitively, this analysis suggests examining the Hamming distance, which reports the number of disagreeing indices:
\begin{equation}
    |P_1 - P_2|_\text{Ham} = \sum_{i \in [N]} \mathbbm{1}_{P_1[i] \neq P_2[i]}. \label{eq:ham_dist}
\end{equation}
However, this undercounts the number of CNOTs needed. On indices where $P_1[i] \neq P_2[i]$ but neither is $I$, we see that $P_1[i]$ and $P_2[i]$ will each originate (control) a CNOT, but Eq.~\ref{eq:ham_dist} only increments by 1 on index $i$. To adjust this undercounting, we define the corrected CNOT distance as
\begin{equation}
\begin{split}
    |P_1 - P_2|_\text{CNOT} &:= |P_1 - P_2|_\text{Ham} + \sum_{i \in [N]} \mathbbm{1}_{I \ \neq P_1[i] \neq P_2[i] \neq I} \\
    &= \sum_{i \in [N]} \mathbbm{1}_{P_1[i] \neq P_2[i]}(1 + \mathbbm{1}_{I \not\in \{P_1[i], P_2[i]\}}).
    \label{eq:cnot_distance}
\end{split}
\end{equation}
This distance, which is symmetric in its arguments, reports the number of CNOTs needed to implement the transition zone from $\exp{(-i P_1 t)}$ to $\exp{(-i P_2 t)}$, after performing gate cancellation.

Given this distance, we could in principle find the ordering with minimum CNOT cost by summing the CNOT cost along each of the possible $\pi$ permutation functions. However, this approach would be impractical, taking $O(k!)$ factorial time. Instead, we can formulate our objective as a TSP problem, in which we are given a weighted graph and asked to find the shortest cycle visiting each vertex once.

This bijects clearly to our setting---we simply define a complete graph $G$ on $k$ vertices and set the distance between vertices $i$ and $j$ as the CNOT cost $|P_i - P_j|_\text{CNOT}$. Thus, TSP gives a permutation that cycles through $\{P_1, ..., P_k\}$ with minimal CNOT cost. To be precise, we actually seek the shortest Hamiltonian path; in other words, we want to visit each Pauli string once, without returning to the start. We can do so by creating a virtual node connected to every other node with cost 0, as described by the \textit{computer wiring} bijection in \cite{lenstra1975some}. Algorithm~\ref{alg:cnot_gate_cancellation} shows pseudocode for the full gate cancellation subroutine.

However, TSP is NP-hard. While it can be solved in $O(k^2 2^k)$ time by the Bellman-Held-Karp dynamic algorithm \cite{bellman1961dynamic, held1962dynamic}---an improvement over the $O(k!)$ brute-force solution---it is still exponentially expensive to solve in general. It would be impractical to invest exponential classical resources to optimize a polynomially sized quantum circuit. Fortunately, there are good heuristics and approximation algorithms for TSP, which we discuss next.

\begin{algorithm}[H]
\SetAlgoLined \SetKwData{FinalOrdering}{finalOrdering}
\KwIn{Weighted Pauli strings, unordered: $\{P_1, ..., P_k\}$}
\KwResult{Permutation approximately maximizing CNOT cancellation: $[P_{\pi(1)}, ..., P_{\pi(k)}]$}
 \FinalOrdering$\leftarrow$ $[ \: ]$\;
    $G \leftarrow $ undirected complete graph with $k + 1$ nodes\;
    Set $G[k+1, *] = 0$\;
    \For{$i, j \in [k] \times [k]$}{
          Dist[$i, j$] = $|P_i - P_j|_\text{CNOT}$\;
    }
    Solve TSP on $G$\;
    Delete $(k + 1)$th vertex in TSP route to break cycle\;
    Append Hamiltonian path to \FinalOrdering\;
 \KwRet{\FinalOrdering}
 \caption{CNOT Gate Cancellation} \label{alg:cnot_gate_cancellation}
\end{algorithm}

\subsection{TSP Approximation} \label{sec:tsp_approx}
Solving TSP in the most general setting is NP-hard. Moreover, no polynomial time algorithms exists that is guaranteed to approximate it to a constant ratio \cite{sahni1976p}. For the case of \textit{metric graphs}, however, TSP can be efficiently 1.5-approximated via Christofides' algorithm \cite{christofides1976worst}, meaning that the approximation will return an ordering that requires at most 1.5x as many CNOTs as the optimal ordering does. Moreover, Christofides' algorithm is fast, running in $O(k^3)$ time as originally proposed and later improved to $\tilde{O}(k^2)$ \cite{chekuri2017approximating, chekuri2018fast, google2018christofides}. In practice, Christofides' algorithm, as well as other heuristics that do not have rigorous approximation ratio guarantees, can perform well in practice, often attaining near-optimal solutions \cite{haque2013empirical, genova2017experimental}.

Indeed, the graph defined by the $|P_1 - P_2|_\text{CNOT}$ distance function is a metric graph. We have already seen that it is symmetric in its arguments (i.e., the graph is undirected), so we need only show now that it satisfies the triangle inequality:
\begin{equation}
|P_1 - P_2|_\text{CNOT} + |P_2 - P_3|_\text{CNOT} - |P_1 - P_3|_\text{CNOT} \geq 0 \nonumber .
\end{equation} We prove this below.
\begin{equation}
\begin{split}
|P_1 - P_2|_\text{CNOT} + |P_2 - P_3|_\text{CNOT} - |P_1 - P_3|_\text{CNOT}\\
= \sum_{i \in [N]}
\mathbbm{1}_{P_1[i] \neq P_2[i]}(1 + \mathbbm{1}_{I \not\in \{P_1[i], P_2[i]\}}) \\
+ \mathbbm{1}_{P_2[i] \neq P_3[i]}(1 + \mathbbm{1}_{I \not\in \{P_2[i], P_3[i]\}}) \\
- \mathbbm{1}_{P_1[i] \neq P_3[i]}(1 + \mathbbm{1}_{I \not\in \{P_1[i], P_3[i]\}})
\end{split} \label{eq:big_sum}
\end{equation}
We will prove that each three-term expression in the sum is non-negative for each $i$, so that the full sum must also be non-negative. The third term is 0, -1, or -2.
\begin{itemize}
    \item If it is 0, the three-term expression is already non-negative since the first two terms are non-negative.
    \item If it is -1, then $P_1[i] \neq P_3[i]$, and one of the two is $I$. We must also have $P_1[i] \neq P_2[i]$ or $P_2[i] \neq P_3[i]$, so the first two terms must sum to at least 1. Thus, the three-term expression is non-negative.
    \item If it is -2, then $P_1[i] \neq P_3[i]$, and neither is $I$. Suppose that $P_1[i] = P_2[i]$; then the first term is +2, and the three-term expression is non-negative. Similarly, if $P_2[i] = P_3[i]$, then the second term is +2, and the three-term expression is non-negative. And if $P_1[i] \neq P_2[i] \neq P_3[i]$, then the first two terms are both at least +1, so the three-term expression is non-negative.
\end{itemize}
Thus, we conclude that Eq.~\ref{eq:big_sum} is a sum over non-negative numbers, so the final result is non-negative. This proves that the triangle inequality holds for the CNOT distance, and thus our graph is metric. Therefore, Christofides' algorithm can be used to efficiently attain a 1.5-approximation to the optimal TSP.

However, this approach is not directly compatible with the shortest Hamiltonian path bijection described previously. This bijection required adding a virtual node connected to every other node with zero cost, which would break the triangle inequality. We instead propose to approximate the shortest Hamiltonian path by first approximating TSP and then deleting the most expensive edge to break the cycle. Note that this patch requires two levels of approximation: first approximating the TSP cycle and then approximating the shortest Hamiltonian path based on this cycle. This approach amounts to modifying Algorithm~\ref{alg:cnot_gate_cancellation} by (1) initializing $G$ without the $(k+1)$th node, (2) approximating TSP instead of solving it exactly, and (3) deleting the most expensive edge in TSP instead of deleting the $(k+1)$th vertex to realize a Hamiltonian path. There may be more direct approaches to approximating the shortest Hamiltonian path, but TSP is better studied and has more well-known approximation algorithm results.

\subsection{Advantage Over Lexicographic Ordering}

\textit{Lexicographic} ordering does often lead to good gate cancellation properties and has the advantage of being extremely fast to compute, for example by a linear-time string sort. However, it does not achieve the optimality of TSP. As an example, consider the 4-qubit Hamiltonian with lexicographically ordered strings $[XXXX$, $XXYY$, $XYXY$, $XYYX$, $YXXY$, $YXYX$, $YYXX$, $YYYY]$. These eight commuting strings arise in the Jordan-Wigner encoding of $H_{pqrs}$ for molecules, so this example is ubiquitous \cite{gokhale2019minimizing, whitfield2011simulation}. Applying Eq.~\ref{eq:cnot_upper_bound}, we see that $8 \times 2 \times 4 = 64$ CNOTs are needed prior to gate cancellation. After CNOT cancellation, summing Eq.~\ref{eq:cnot_distance} along the lexicographic order gives a total of 40 CNOTs. Now consider the TSP order $[XXXX$, $XXYY$, $XYXY$, $XYYX$, $YXYX$, $YXXY$, $YYXX$, $YYYY]$, which flips the third and fourth terms from the lexicographic order. Under TSP, we are able to execute the dynamics with just 36 CNOTs. In summary, unoptimized to \textit{lexicographic} to TSP have CNOT costs of $64 \to 40 \to 36$.

In certain cases, TSP can have an even greater factor of improvement over \textit{lexicographic} ordering. For example, consider the nine Pauli strings below, specially annotated and monospaced for pattern readability.

\begin{enumerate}
    \item \texttt{XX|XXXXXXXX} 
    \item \texttt{XY|XXXZZZZZ}
    \item \texttt{XZ|XXXXXXXZ}
    \item \texttt{YX|XXZZZZZZ}
    \item \texttt{YY|XXXXXXZZ}
    \item \texttt{YZ|XZZZZZZZ}
    \item \texttt{ZX|XXXXXZZZ}
    \item \texttt{ZY|ZZZZZZZZ}
    \item \texttt{ZZ|XXXXZZZZ}
    
\end{enumerate}

Without gate cancellation, $9 \times 10 \times 2 = 180$ CNOTs are required. Under the given lexicographic order above, gate cancellation yields 112 CNOTs. However, with reordering into the TSP route, only 62 CNOTs are needed. The TSP route is as follows.
\begin{enumerate}[(i)]
    \item \texttt{XX|XXXXXXXX} 
    \item \texttt{XZ|XXXXXXXZ}
    \item \texttt{YY|XXXXXXZZ}
    \item \texttt{ZX|XXXXXZZZ}
    \item \texttt{ZZ|XXXXZZZZ}
    \item \texttt{XY|XXXZZZZZ}
    \item \texttt{YX|XXZZZZZZ}
    \item \texttt{YZ|XZZZZZZZ}
    \item \texttt{ZY|ZZZZZZZZ}
\end{enumerate}


In fact, this is a special case of an \textit{asymptotic advantage} in gate cancellation that TSP can achieve over \textit{lexicographic} ordering. Consider, a set of $N+1$ Pauli strings, where each Pauli string is the concatenation of two of Pauli substrings, $P_\text{index}|P_{XZ(i)}$. $P_{XZ(i)}$, defined for $i \in [N]$, has width of $N$ and consists of $(N - i)$ $X$'s followed by $i$ $Z$'s. Note that $P_{XZ(i)}$ and $P_{XZ(i + \lfloor N/2 \rfloor \mod N)}$ differ by a Hamming distance of $\sim N/2$ and therefore have a CNOT distance of $O(N)$. However, $P_{XZ(i)}$ and $P_{XZ(i) + 1}$ have a CNOT distance of $O(1)$ since only one index changes. We denote these two sequences as \textit{unfavorable} and \textit{favorable}, respectively.

An adversarial strategy follows from these sequences. We prefix the $P_{XZ(i)}$'s with $P_\text{index}$, of width $O(\log{N})$. We assign these indices so that lexicographically sorting the Pauli strings arranges the $P_{XZ(i)}$ in the \textit{unfavorable} sequence where each adjacent pair has $O(N)$ CNOT distance. In this sequence, the total CNOT distance is $O(N^2)$, which has no asymptotic improvement over the no-gate-cancellation baseline. The first list above is lexicographically ordered into the unfavorable sequence---notice that each adjacent $P_{XZ(i)}$ pair has a Hamming distance of 4 or 5.

By contrast, TSP would disregard the $P_{index}$ qubits and instead optimizes the CNOT distances by choosing the favorable sequence. In this ordering, there are a total of $O(N \log{N})$ CNOTs along the $P_{index}$ indices and $O(N \times 1) = O(N)$ CNOTs along the $P_{XZ(i)}$ indices, amounting to $O(N \log{N})$ CNOTs via TSP. The second list above (with Roman numeral headings) shows the TSP route---note that each adjacent $P_{XZ(i)}$ pair has a Hamming distance of just 1. The bottleneck in CNOT cost therefore originates from the $P_\text{index}$ substrings, which have only logarithmic width.

While this particular example is pathological, it demonstrates scenarios where \textit{lexicographic} ordering is asymptotically identical to no-gate cancellation, but TSP achieves an asymptotic advantage. This advantage is from $O(N^2)$ to $O(N \log{N})$, which is a significant improvement. For future work, we will evaluate empirical improvements in terms of the gains both from TSP over lexicographic and from approximate/heuristic TSP over \textit{lexicographic}.

\subsection{Extension for Future Cost Functions and Optimizations} \label{subsec:tsp_extensions}

The TSP bijection in Algorithm~\ref{alg:cnot_gate_cancellation} is flexible in that it can be extended to other cost functions, i.e, new $\text{Dist}[i, j]$ assignments. One natural extension would be to account for single-qubit gate cancellations in addition to CNOT cancellations. For example, the cancellation between two Pauli $X$ characters may be preferable to the cancellation between two Pauli $Z$ characters, since the latter cancels out Hadamard gates in addition to a CNOT pair. The cost function could also account for hardware topological constraints or cancellations between non-nearest-neighbor Pauli strings. However, these cost functions might not be metric, which could hamper the approximability of TSP.

We could also explore different circuit architectures for the $\exp{-i P_{\pi(j)} t}$ terms. For example, the ancilla could be removed, and instead the $R_z$ could target the last non-$I$ qubit, as in Figures~\ref{fig:Center_Original} (\textit{star}) and \ref{fig:Ladder_Original} (\textit{ladder}). In the corresponding \textit{star} architecture, we would need to prepartition the Pauli strings into subsets that share common last non-$I$ characters, to ensure that the CNOTs all cancel out. This strategy is analogous to a lexicographic sort on the last non-$I$ character (rather than a lexicographic sort across all indices).


\subsection{Benchmarking Results of TSP for Gate Cancellation}

\begin{figure}[h!]
  \centering
    \includegraphics[width=0.9\columnwidth]{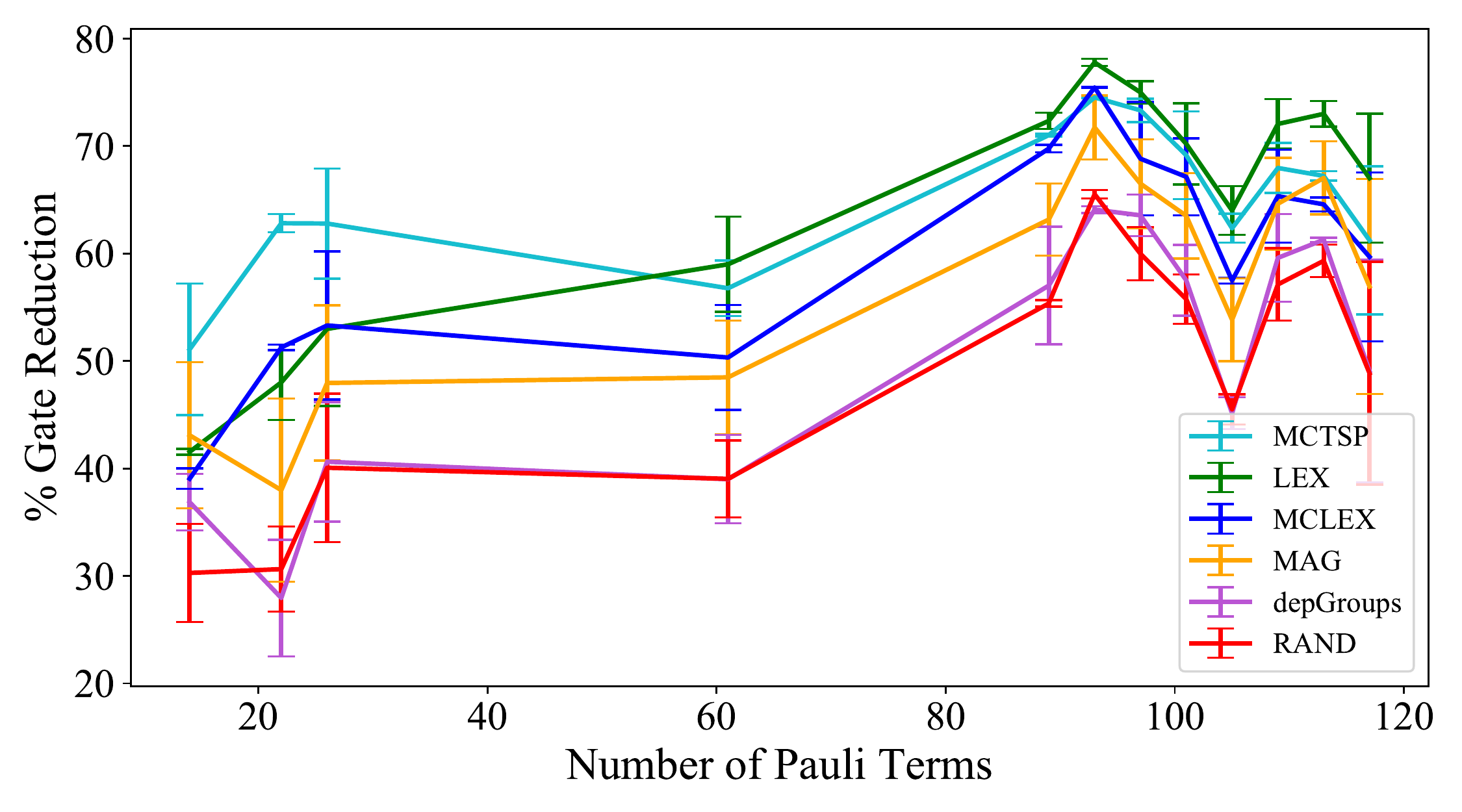}
    \caption{Gate reduction as a function of the number of Pauli terms within the simulated Hamiltonian. As the number of terms increases, the \textit{lexicographic} ordering surpasses the gate cancellation of the \textit{max-commute-tsp} ordering. Although \textit{lexicographic} simulation circuits may have greater gate cancellation, their poor process fidelity will require a larger Trotter number --- resulting in longer circuits overall.}
    \label{fig:gate_reduction_plot}
\end{figure}

Here we show the gate cancellation percentages of different ordering methods in Fig~\ref{fig:gate_reduction_plot}, over different benchmarking molecules with different numbers of Pauli terms. 

We can observe that \textit{lexicographic} produces the greatest circuits cancellation percentage, while \textit{max-commute-tsp} produces the second-best cancellation percentage. This is because \textit{max-commute-tsp} only optimizes the ordering of Pauli terms via TSP within commuting cliques, the amount of gate cancellation in \textit{max-commute-tsp} circuits is necessarily less. However, we point out that a TSP ordering over the entire Hamiltonian would quickly grow intractable, and it would not necessarily have a high process fidelity. In fact, to make up for its poor process fidelity, a \textit{lexicographic} ordering must increase its Trotter number, concatenating $r$ copies of the circuit together: quickly erasing any gate savings seen at the $r=1$ level.

Figure~\ref{fig:gate_reduction_plot} also shows the advantage that a  \textit{max-commute-tsp} ordering has over \textit{max-commute-lex}, which lexicographically orders terms within the cliques. Because the number of terms within each clique is relatively small compared to the entire Hamiltonian, finding an ordering within a clique is similar to the few Pauli term region of Figure~\ref{fig:gate_reduction_plot}. In this regime, the TSP heuristic is able to produce much more gate cancellations than any other ordering strategy. The \textit{max-commute-tsp} ordering carries this advantage through to the many Pauli term regime where it outperforms the \textit{max-commute-lex} ordering while maintaining a high process fidelity. 
\section{Combined Benchmarking}
\label{sec:combined_benchmarking}

Since the goal of this work is to examine compilations which mitigate both physical and algorithmic errors, we simulate all 156 benchmark Hamiltonians under noise models and execute 6 small benchmarks on ion trap quantum computers to combine the effects of optimizations targeting both sources of error. We simulated each of the benchmark Hamiltonians under a depolarizing error model using the \textit{lexicographic}, \textit{magnitude}, and \textit{max-commute-tsp} orderings with three different initial states and four separate error rates. 

\subsection{Methodology}
The combined effects of algorithmic and physical errors are captured by noisy simulations using a depolarizing error model which adds a noise channel on the two-qubit gates with varying error rates. The depolarizing channel is defined as \cite{nielsen2002quantum}:
\begin{equation}
    \mathcal{E} (\rho) = \frac{pI}{2} + (1 - p)\rho.\nonumber
\end{equation}
This means that each entangling gate has probability $p$ of depolarizing its control and target qubits (in other words, replacing the qubits with the completely mixed state $I/2$), and probability $1 - p$ of leaving the qubits untouched.

Recent work has demonstrated the impact that the choice of initial state has on Digital Quantum Simulation~\cite{chen2020quantum}. Therefore, we selected three initial states to cover varying degrees of entanglement and superposition.

\begin{figure}[h]
  \centering
    \includegraphics[width=0.9\columnwidth]{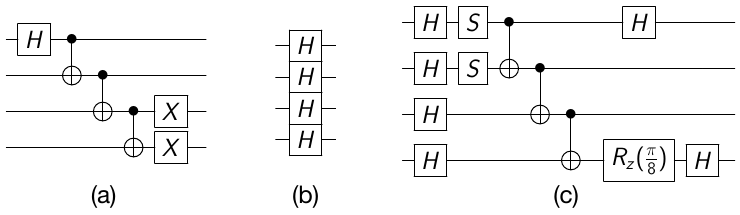}
    \caption{Initial states used in the noisy Digital Quantum Simulations. The three states capture different notions of initial state complexity. (a) creates a state spanning only two basis states which are entangled. (b) produces an equal superposition of all basis states but contains no entanglement. (c) is a combination of the two, spanning more than two basis states and exhibiting entanglement.}
    \label{fig:initial_states}
\end{figure}

The Hellinger infidelity \cite{harper2020efficient, viacheslav2005operational, qiskit2020hellinger} is used to measure the ability of the \textit{lexicographic}, \textit{magnitude}, and \textit{max-commute-tsp} orderings to evolve the initial states under each Hamiltonian with noisy quantum gates. The Hellinger infidelity is defined as $1 - H_D(P,Q)$, where $H_D(P,Q)$ is the Hellinger distance, defined for two probability distributions $P$ and $Q$ as
\begin{equation}
    H_D (P,Q) = \frac{1}{\sqrt{2}} ||\sqrt{P}-\sqrt{Q}||_2.\nonumber
\end{equation}

We use this metric for the noisy simulations and trapped ion experiments, instead of the process fidelity, because it allows us to use the measurement results of noisy, shot-based program executions. For these benchmarks, the Hellinger infidelity is computed by first evolving an initial state from Figure~\ref{fig:initial_states} with an DQS circuit generated via a \textit{lexicographic}, \textit{magnitude}, or \textit{max-commute-tsp} term ordering. Then the resulting quantum state is sampled to produce a probability distribution $Q$ which is compared with the expected distribution $P$ obtained via exact simulation. All of the code used to generate and test the benchmarks is available online in a Github repository~\cite{anonymous2020github}.

\subsection{Benchmarking Results and Analysis}

\begin{figure}[t]
    \centering
    \includegraphics[width=1.0\columnwidth]{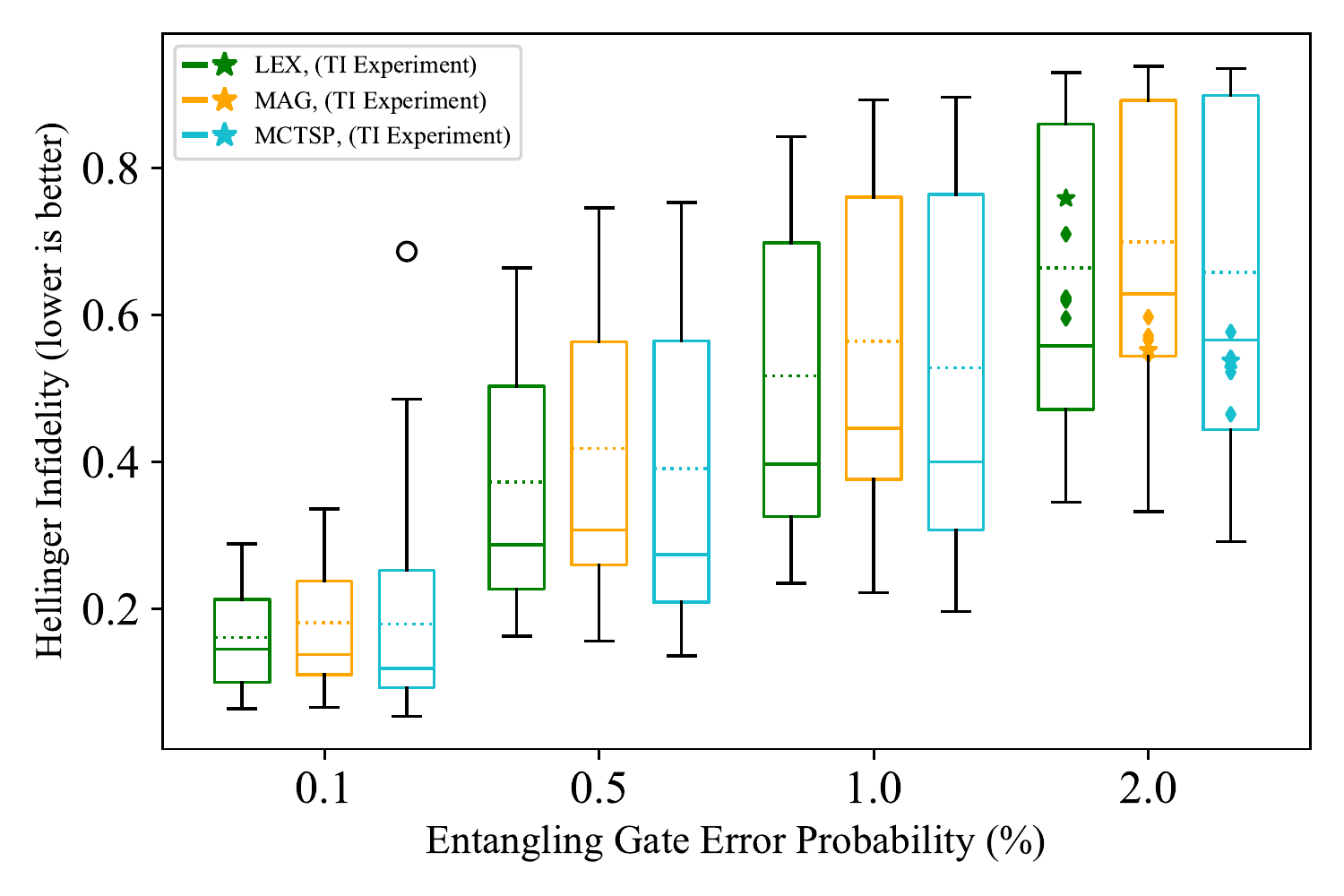}
    \caption{Distribution of Hellinger infidelities after evolving the entangled initial state ($\frac{1}{\sqrt{2}}(\ket{0011} + \ket{1100})$). The dotted (solid) lines within each box plot denote the mean (median) of the distribution. The edges of the boxes represent the upper and lower quartiles of the data while the whiskers show the maximum and minimum values (excluding the outliers which are denoted by the open circles). The stars and diamonds indicate experimental results obatained via a 7- and 11-qubit ion trap processor, respectively~\cite{alderete2020quantum, wright2019benchmarking}.}
    \label{fig:ghz_init}
\end{figure}

We first examined a simple initialization state: $\frac{1}{\sqrt{2}}(\ket{0011}$ + $\ket{1100})$ (Figure~\ref{fig:initial_states}a) under noise model simulations. The whisker boxes in Figure~\ref{fig:ghz_init} shows the measured distribution of Hellinger infidelities. The \textit{max-commute-tsp} ordering produces DQS circuits with lower infidelity on average and also attains a minimum infidelity that is 1.2\%, 2.4\%, 3.6\%, 6.7\% lower than the minimum achieved by either the \textit{lexicographic} or \textit{magnitude} orderings for each of the 0.1\%, 0.5\%, 1.0\%, 2.0\% error rates, respectively. 

\begin{table}[]
\begin{tabular}{cccc}
\multicolumn{1}{l}{}                    & \multicolumn{1}{l}{}                   & \multicolumn{1}{l}{}               & \multicolumn{1}{l}{}                       \\ \hline
\multicolumn{1}{|c|}{}          & \multicolumn{1}{c|}{Process Fid.(\%)} &
\multicolumn{1}{c|}{2-qubit gates} & \multicolumn{1}{c|}{Hellinger Infid.(\%)} \\
\multicolumn{1}{|c|}{}          & \multicolumn{1}{c|}{\textit{lex}, \textit{mag}, \textit{mctsp}} &
\multicolumn{1}{c|}{\textit{lex}, \textit{mag}, \textit{mctsp}} & \multicolumn{1}{c|}{\textit{lex}, \textit{mag}, \textit{mctsp}} \\
\hline
\multicolumn{1}{|c|}{\textit{$C_2H_4$}} & \multicolumn{1}{c|}{29.2, 100, 100}    & \multicolumn{1}{c|}{55, 49, 41}    & \multicolumn{1}{c|}{75.9, 55.2, \textbf{53.8}}      \\
\multicolumn{1}{|c|}{\textit{$Cl_2$}}   & \multicolumn{1}{c|}{28.2, 100, 100}    & \multicolumn{1}{c|}{47, 53, 37}    & \multicolumn{1}{c|}{62.2, 56.6, \textbf{54.2}}      \\
\multicolumn{1}{|c|}{\textit{$C_2H_2$}} & \multicolumn{1}{c|}{28.1, 100, 100}    & \multicolumn{1}{c|}{47, 53, 37}    & \multicolumn{1}{c|}{62.4, \textbf{57.2}, 57.7}      \\
\multicolumn{1}{|c|}{\textit{$F_2$}}    & \multicolumn{1}{c|}{71.3, 100, 100}    & \multicolumn{1}{c|}{47, 53, 37}    & \multicolumn{1}{c|}{71.0, 56.7, \textbf{52.2}}      \\
\multicolumn{1}{|c|}{\textit{$N_2$}}    & \multicolumn{1}{c|}{43.1, 100, 100}    & \multicolumn{1}{c|}{47, 53, 37}    & \multicolumn{1}{c|}{61.9, 54.6, \textbf{54.1}}      \\
\multicolumn{1}{|c|}{\textit{$O_2$}}    & \multicolumn{1}{c|}{22.7, 100, 100}    & \multicolumn{1}{c|}{41, 41, 27}    & \multicolumn{1}{c|}{59.6, 59.7, \textbf{46.5}}      \\ \hline
\end{tabular}
\caption{Process fidelities, gate counts, and Hellinger infidelities for the benchmarks tested on ion trap QPUs. Each DQS circuit utilized 4 data qubits and 1 ancilla qubit, and was prepared in the entangling initial state (Fig.~\ref{fig:initial_states}a). The $C_2H_4$ benchmark was run on a 7-qubit processor with 4000 shots~\cite{alderete2020quantum}. The remainder were evaluated on an 11-qubit processor from IonQ with 1000 shots each~\cite{wright2019benchmarking}.}
\label{tab:trappedion_experiment}
\end{table} 
We also evaluated the performance of the ordering strategies using trapped ion quantum computers across six benchmark molecules (Table~\ref{tab:trappedion_experiment}) for this initial state. The ethene ($C_2H_4$) benchmark was run on a 7-qubit device~\cite{alderete2020quantum} while the remainder utilized an IonQ device with 11-qubits~\cite{wright2019benchmarking}. Both processors have entangling gate errors near 2\%. The experimental results are in shown in Figure~\ref{fig:ghz_init} as the stars (7-qubit) and diamonds (11-qubit), and Table~\ref{tab:trappedion_experiment} contains their process fidelities, gate counts, and Hellinger infidelities. The hardware experiments confirm the results of the noisy-simulations and also highlight the impact of algorithmic errors -- even for current NISQ processors. For the $Cl_2$, $C_2H_2$, $F_2$, and $N_2$ benchmark molecules the \textit{magnitude} ordering, which produced deeper quantum circuits with higher process fidelity, attained lower infidelities than the \textit{lexicographic} ordering, which produced shorter circuits with lower process fidelity. Interestingly, both strategies produced circuits of equal depth for the $O_2$ benchmark and achieved similar Hellinger infidelities despite a large difference in process fidelity. This may be due to the specific choice of initial state since the process fidelity is a measure of accuracy over all initial states while the reported Hellinger infidelity is measured with respect to a single initial state.

\begin{figure}[t]
    \centering
    \includegraphics[width=0.95\columnwidth]{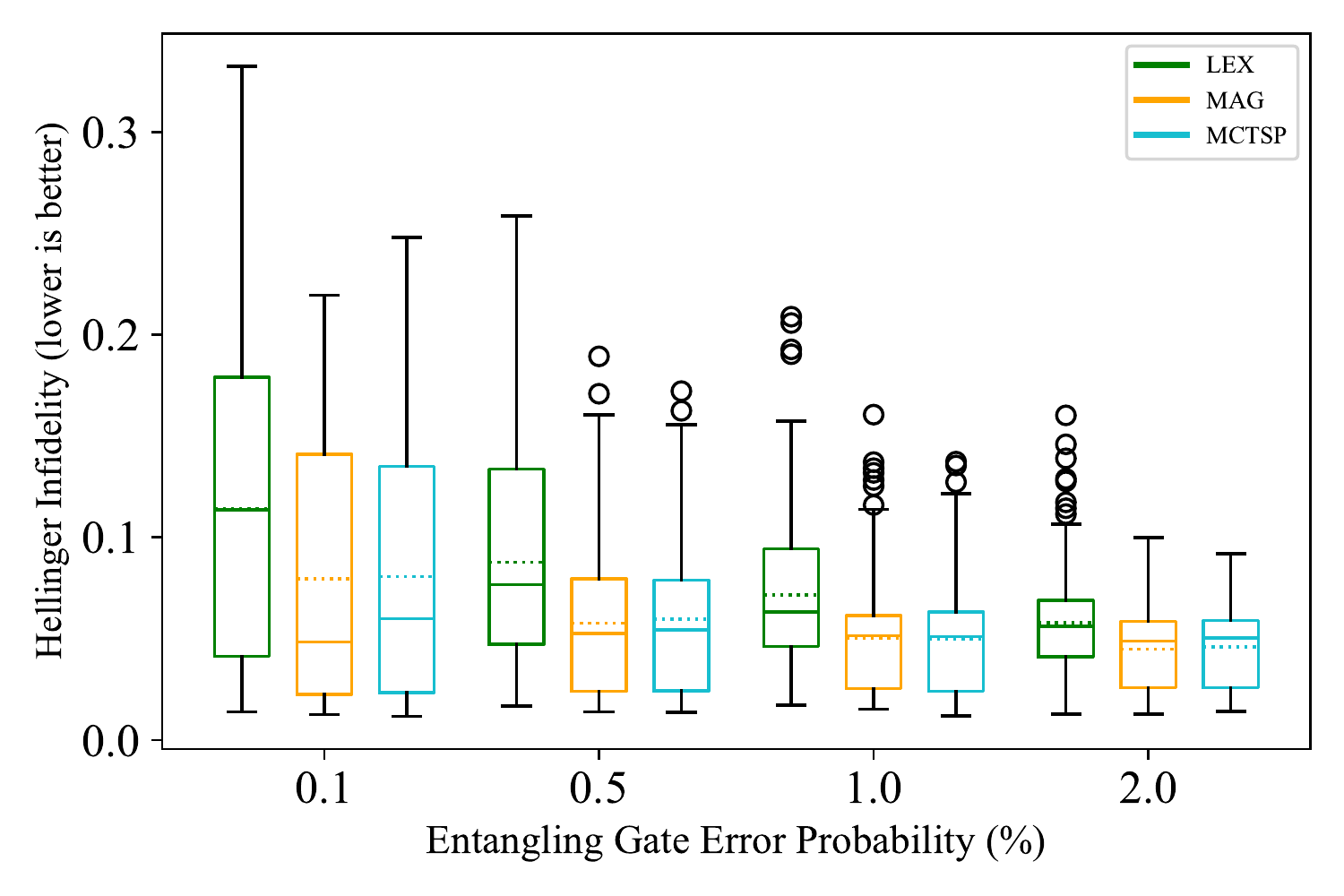}
    \caption{Distribution of Hellinger infidelities across all benchmarks, this time the initial state is an equal superposition state over all of the computational basis states ($\frac{1}{\sqrt{N}}\sum_i^N\ket{i}$).}
    \label{fig:superposition_init}
\end{figure}

In Figure~\ref{fig:superposition_init} we performed the same evaluation using an equal superposition initial state (Figure~\ref{fig:initial_states}b). Unlike the entangled initial state, which is a superposition over two basis states, the Figure~\ref{fig:initial_states}b initial state requires that each of the $2^N$ basis states is accurately evolved to ensure high accuracy overall. In the low-error regime, \textit{lexicographic} has high infidelity while \textit{magnitude} and \textit{max-commute-tsp} have better performance. While all three orderings achieve comparable minimum infidelities, on average, the range of Hellinger infidelities produced by \textit{max-commute-tsp} is 20.3\% smaller than the other strategies.
Interestingly, the simulations display a decrease in infidelity as the error rate increases. This is because the final state of the ideal evolution is also a nearly equal superposition state, and so as the gate error rate increases the final output of the \textit{lexicographic}, \textit{magnitude}, and \textit{max-commute-tsp} circuits more and more resembles a purely noisy state in equal superposition. Therefore, the differences between the ordering strategies in Figure~\ref{fig:superposition_init} can be attributed to their average process fidelities.

\begin{figure}[h!]
    \centering
    \includegraphics[width=0.95\columnwidth]{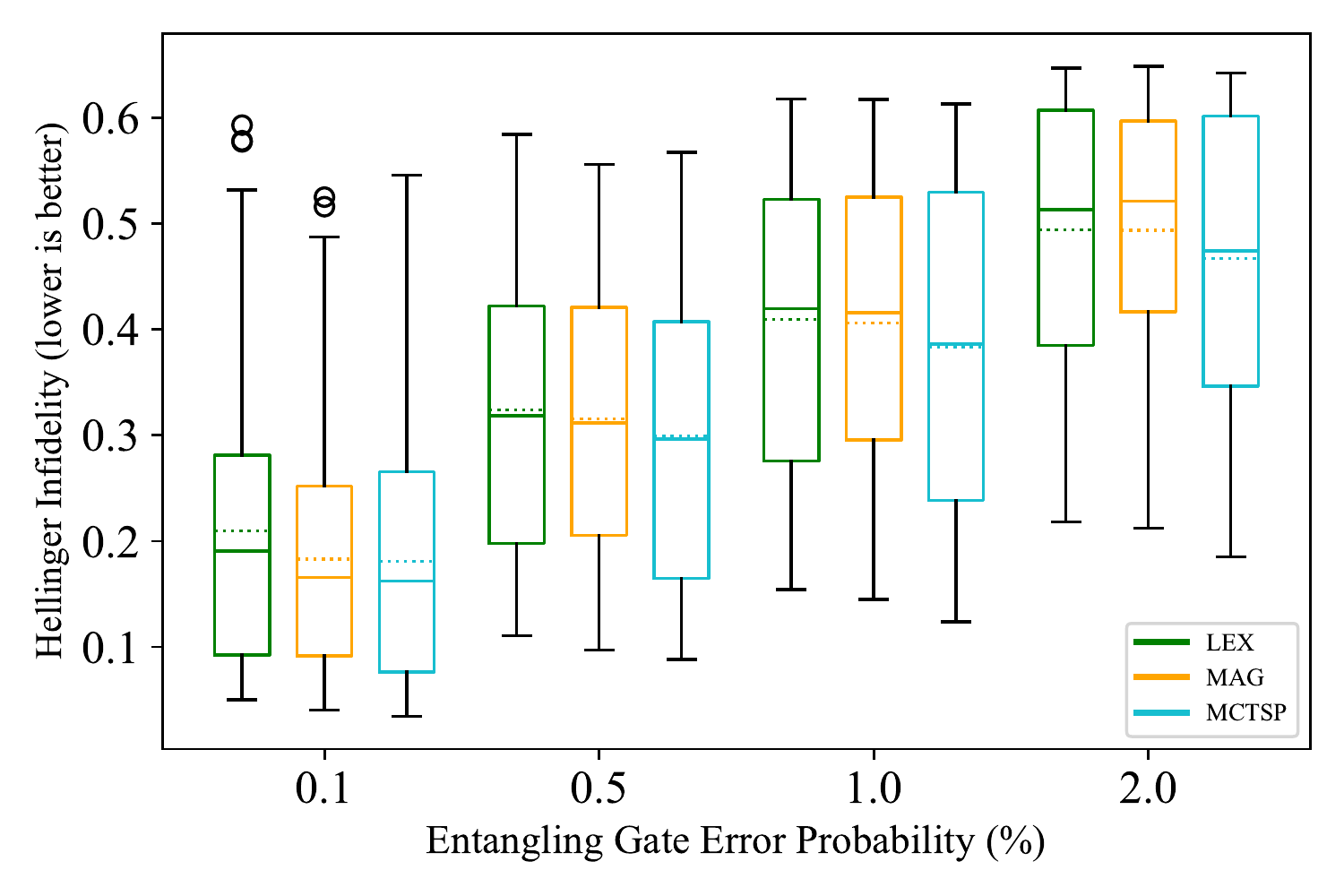}
    \caption{Distribution of Hellinger infidelities after evolving the Fig.~\ref{fig:initial_states}c initial state. The distribution of circuits produced by the \textit{max-commute-tsp} ordering have consistently lower infidelities than the other strategies.}
    \label{fig:ave_complex}
\end{figure}

The final benchmarks were initialized with the circuit in Figure~\ref{fig:initial_states}c. This is a more complex initial state which contains multiple entangled basis states, different phases, and different amplitudes. The combined impact of algorithmic and physical errors is shown in Figure~\ref{fig:ave_complex}. In the low-error regime all three compilations perform similarly, but as the probability of errors increases, the improved performance of \textit{max-commute-tsp} becomes clear. Similar to Figure~\ref{fig:ghz_init}, \textit{max-commute-tsp} attains minimum infidelities which are 1.1\%, 1.6\%, 2.6\%, 3.0\% lower (at each of the respective error levels) than those obtained by \textit{lexicographic} or \textit{magnitude}. For some Hamiltonians, this improvement over \textit{lexicographic} and \textit{magnitude} reaches 58\% and 11\%, respectively.
\section{Appendix}
Lemma 1: Binomial theorem for matrices: For arbitrary Hermitian matrices A and B of the same size, if $[A, B] = 0$, then we can binomially expand $(A+B)^{n}$
\begin{proof}
In general, expanding out $(A + B)^{n}$ would result in a summation of the permutation of the string with A and B, or namely, $\{A, B\}^{\bigotimes n}$. Since A and B commute with each other, we can freely arrange the orders of A and B in each AB strings, and group them together just like the binomial series for numbers.
\begin{equation}
    (A + B)^{n} = \sum_{l=0}^{n}\binom{n}{l}A^{n - l}B^{l} \nonumber
\end{equation}
\end{proof}

Corollary 1: In a multiplication series of matrix $A_{1}A_{2}...A_{n}$, if [$A_{j}$, $A_{k}$] = 0 $\forall j, k \in \{1, 2,..., n\}$, we can binomially expand the series as $(A_{1}A_{2}...A_{n-1} + A_{n})^k$
\begin{proof}
By induction. Take Theorem 1 as base case, and induct on the newly added $A_{n}$ term.
\end{proof}

\subsubsection{Theorem 1 Proof}
\begin{proof}

Using Taylor expansion and Lemma 1, we have
\begin{align}
e^{A+B}&=\sum_{k=0}^{\infty}\frac{1}{k!}(A+B)^{k}\nonumber\\
&=\sum_{k=0}^{\infty}\frac{1}{k!}\sum_{l=0}^{k}\binom{k}{l}A^{k-l}B^{l}\nonumber\\
&=\sum_{k=0}^{\infty}\sum_{l=0}^{k}\frac{1}{k!}\frac{k!}{l!(k-l)!}A^{k-l}B^{l}.\nonumber
\end{align}
Since $l$ is bounded by $k$, as $k \to \infty$, $l$ will sum up to $\infty$; therefore we can swap $k$ and $l$ as
\begin{align}
e^{A+B}&=\sum_{k=l}^{\infty}\sum_{l=0}^{\infty}\frac{1}{l!(k-l)!}A^{k-l}B^{l}.\nonumber
\end{align}
Let $k - l = m$, we have
\begin{align}
e^{A+B}&=\sum_{m=0}^{\infty}\sum_{l=0}^{\infty}\frac{1}{l!m!}A^{m}B^{l}\nonumber\\
&=e^{A}e^{B} .\nonumber
\end{align}
\end{proof}

\subsubsection{Theorem 2 Proof}
\begin{proof}
By induction. For simplicity, we substitute $-iH_{1}t = h_{1}$, $-iH_{2}t = h_{2} ,..., -iH_{M}t = h_{M}$. The commutation relations still hold since $i$ and $t$ are just coefficients.\\

Base case: application of Theorem 1.

Induction hypothesis: Suppose the theorem holds for all values of  $M$, up to some number $P$, $P \geq$ 3.
\\
Induction steps: Let $M = P + 1$.\\
Denote $\sum_{n=1}^{P}h_{n}$ as $H_{BEG}$.
By induction hypothesis, we have
\begin{equation}
    e^{H_{BEG}} = e^{h_{1}}e^{h_{2}}...e^{h_{P}}. \nonumber
\end{equation}
Since $h_{P+1}$ commutes with all the other $h_{n}$ terms, from Corollary 1 we have
\begin{equation}
    e^{H_{BEG} + h_{P+1}} = \sum_{k=0}^{\infty}\frac{1}{k!}(H_{BEG}+h_{P+1})^{k}. \nonumber
\end{equation}
Using the same technique from Theorem 2's proof by letting $H_{BEG}$ = A and $h_{P+1}$ = B, we have
\begin{equation}
    e^{H_{BEG} + h_{P+1}} = e^{H_{BEG}}e^{h_{P+1}}. \nonumber
\end{equation}
Therefore,
\begin{equation}
    e^{\sum_{n}^{P+1}h_{n}} = e^{h_{1}}e^{h_{2}}...e^{h_{P}}e^{h_{P+1}}. \nonumber
\end{equation}
\end{proof}
\section*{Acknowledgments}

The authors would like to thank Nathan Wiebe for the introduction to HS and for helpful discussions on commuting Pauli terms. We thank Pavel Lougovski for initial thoughts and helpful discussions and Yipeng Huang, Xiaoliang Wu and Zain Saleem for helpful comments.

Funding acknowledgments:
P. G. is supported by the Department of Defense (DoD) through the National Defense Science \& Engineering Graduate Fellowship (NDSEG) Program.
P. G., Y. S., and F. C. are funded in part by EPiQC, an NSF Expedition in Computing, under grant CCF-1730449, by the NSF STAQ project under grant NSF Phy-1818914, and by DOE grants DE-SC0020289 and DE-SC0020331.  Y. S. is also funded in part by the NSF QISE-NET fellowship under grant number 1747426.
T. T. and M. M. are also funded by EPiQC, under grant CCF-1730082.
The work of K. G. and M. S. is supported by the U.S. Department of Energy, Office of Science, under contract number DE-AC02-06CH11357.

\newpage

\bibliographystyle{unsrt}
\bibliography{refs}

\end{document}